\documentclass[sigconf]{acmart}
\usepackage[american]{babel}
\usepackage{mathtools} %
\usepackage{booktabs} %
\usepackage{tikz} %

\usepackage{subcaption} %
\usepackage{makecell} %
\usepackage{xcolor} %
\usepackage{systeme} %

  \newcommand{\reviewfix}[1]{}

 \usepackage[suppress]{color-edits}
\addauthor{ak}{blue}

\newcommand{\myV}[1]{{\textit{#1}}}  %
\newcommand{\Vs}[1]{\mbox{\textit{\small #1}}}  %

\def\Snospace~{\S{}}

\renewcommand{\times}{\cdot}

\copyrightyear{2025}
\acmYear{2025}
\setcopyright{cc}
\setcctype{by}
\acmConference[FAccT '25]{The 2025 ACM Conference on Fairness, Accountability, and Transparency}{June 23--26, 2025}{Athens, Greece}
\acmBooktitle{The 2025 ACM Conference on Fairness, Accountability, and Transparency (FAccT '25), June 23--26, 2025, Athens, Greece}\acmDOI{10.1145/3715275.3732172}
\acmISBN{979-8-4007-1482-5/2025/06}

\begin{document}

\title[Auditing for Bias in Ad Delivery Using Inferred Demographic Attributes]{Auditing for Bias in Ad Delivery Using Inferred Demographic Attributes}

\author{Basileal Imana}
\email{imana@princeton.edu}
\orcid{0000-0002-6645-7850}
\affiliation{%
  \institution{Princeton University}
  \city{Princeton}
  \state{New Jersey}
  \country{USA}
  \postcode{08544}
}

\author{Aleksandra Korolova}
\email{korolova@princeton.edu}
\orcid{0000-0001-8237-9058}
\affiliation{%
 \institution{Princeton University}
  \city{Princeton}
  \state{New Jersey}
  \country{USA}
  \postcode{08544}
}

\author{John Heidemann}
\email{johnh@isi.edu}
\orcid{0000-0002-1225-7562}
\affiliation{%
 \institution{USC/Information Sciences Institute}
 \city{Los Angeles}
 \state{California}
 \country{USA}
 \postcode{90292}
}

\begin{abstract}
Auditing social-media algorithms
  has become a focus of public-interest research
  and policymaking
  to ensure their fairness across 
  demographic groups such as race, age, and gender in consequential domains such as the presentation of employment opportunities.
However,
  such demographic attributes are often unavailable to auditors and platforms.
When demographics data is unavailable,
  auditors commonly \emph{infer} them from other available information.
In this work, we study the effects of inference error
 on auditing for bias in one prominent application: \emph{black-box} audit of ad delivery using \emph{paired ads}.  
We show that inference error, if not accounted for,
  causes auditing to falsely miss skew
  that exists.
We then propose a way to mitigate the inference error when
  evaluating skew in ad delivery algorithms.
Our method works by adjusting for expected error
  due to demographic inference,
  and it makes skew detection more sensitive when
  attributes must be inferred.
Because inference is increasingly used for auditing,
  our results provide an important addition
  to the auditing toolbox
  to promote correct audits of ad delivery algorithms for bias.
While the impact of attribute inference on accuracy has been studied in
  other domains,
  our work is the first to consider it 
  for black-box evaluation of ad delivery bias,
  when only aggregate data is available
  to the auditor.
\end{abstract}

\maketitle

\section{Introduction}
Digital ad platforms face increased scrutiny 
  from
  public-interest researchers and regulators
  due to their important role in mediating access
  to information and opportunities.
Through external black-box audits researchers have shown that ad delivery algorithms
  can be biased by demographic attributes
  such as race~\cite{Ali2019a},
  gender~\cite{Imana2021} and
  age~\cite{Levi2022} in consequential and legally protected domains such as employment and housing.
Following these findings, Meta was sued by the U.S.~Department of Justice (DoJ)~\cite{FacebookvsHUDCase},
  and in 2022 reached a settlement to deploy a Variance Reduction System (VRS)
  to reduce bias in delivery of ads for economic opportunities including housing, employment, and credit~\cite{metatech, Timmaraju2023}.
This prominent example shows the importance of holding ad
  platforms accountable through external black-box audits.

The state-of-the-art method for black-box auditing of ad delivery algorithms,
  and a key setting for our work,
  uses \emph{paired ads} that are run targeting the same audience
  and at the same time~\cite{Ali2019a, Ali2019b, Imana2021, Imana2024}. 
Bias is then measured by looking at \emph{relative difference} in delivery
  along a demographic attribute of interest to the auditor.
This setup is the only known methodology that can isolate the role of
  algorithmic bias in ad delivery from confounding factors,
  such as market forces and temporal effects (we discuss this setup in \autoref{sec:paired_ads_method}).

\begin{figure*}
  \centering
  \includegraphics[width=0.9\linewidth]{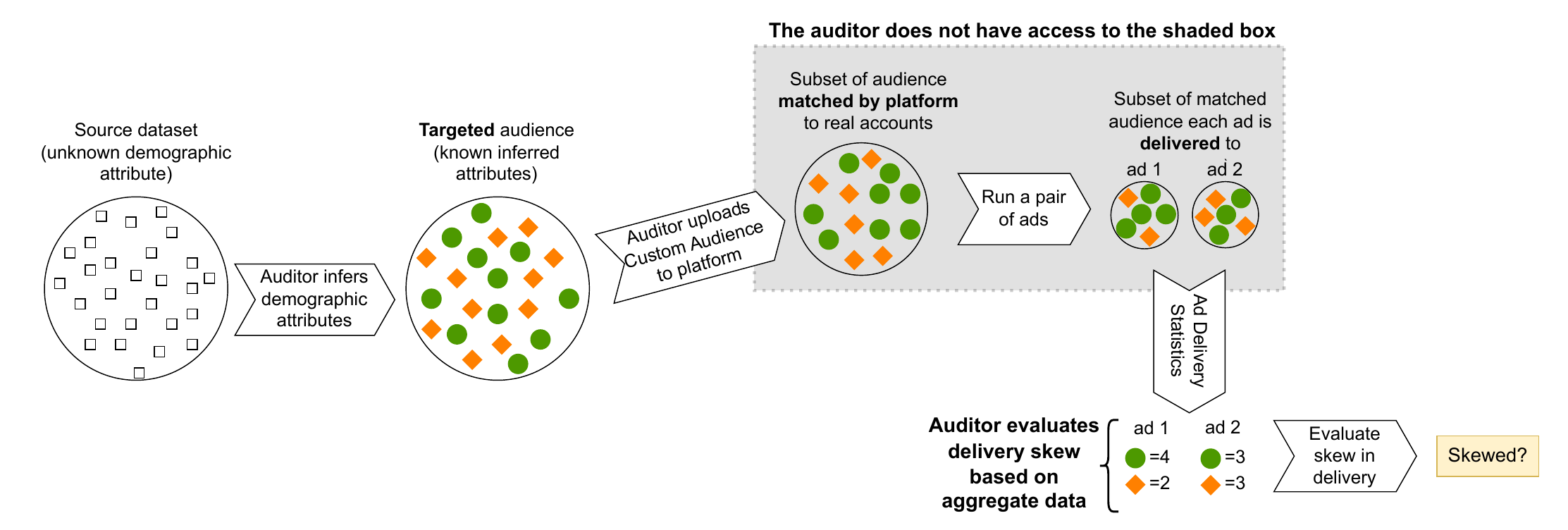}
  \caption{Decoupling between attribute inference step and evaluation of skew in black-box auditing of ad delivery algorithms. Only aggregate size of demographic groups (no individual-level data) is available at the time of skew evaluation.}
  \label{fig:decoupling}
\end{figure*}

A key challenge to applying the paired-ads methodology to
  auditing ad delivery and to expanding such audits to other protected attributes is 
  unavailability of demographic attributes of
  users~\cite{Andrus2021, Bogen2020, Bogen2024}.
Auditors have tackled this challenge
  by using public voter lists from U.S.~states that contain these attributes~\cite{Ali2019a, Ali2019b, Imana2021}.
Some platforms, on the other hand, ask
   users to voluntarily self-identify to conduct internal audits~\cite{LinkedInSelfID},
  but this data is not available for external auditors
  (we summarize such approaches in \autoref{sec:related_work_collecting}).
Another technique that both auditors and platforms
  use is inferring demographic attributes from other available information.
For example, Bayesian Improved Surname Geocoding (BISG), is commonly used
  to infer race from name and location~\cite{Elliott2009, Bogen2024}
  (we expand on these existing approaches in \autoref{sec:related_work_inferring}).
Attribute inference 
  is employed widely across various domains
  as part of 
  assessing demographic disparities and enforcing civil rights laws~\cite{Adjaye2014, Baines2014, Zhang2018, Elzayn2023, Bogen2024},
  despite their known misclassification of a
  significant proportion of individuals~\cite{Elliott2009}.
While approaches to correct inference error have been developed~\cite{Zhu2022WeakPA, Elzayn2023, Lu2024,Mccartan2024},
  they do not directly apply
  in our setting of black-box audits of ad delivery algorithms
  due to two constraints.
First, 
  correction methods assume inferred attributes 
   or inference probabilities of individuals
  are directly accessible
  \emph{at the time of model evaluation}.
Unfortunately ad platforms report only \emph{aggregate} data about ad recipients
  and not per-individual data.
We illustrate this challenge in \autoref{fig:decoupling}:
information about individuals is reserved for the platform (inside the shaded box),
  while auditors see only the proposed targeted audience (left circle)
  and aggregate results (below the shaded box).
Second, prior corrections for inference error
  focus on evaluating group fairness metrics,
  such as Demographic Parity,
  and these do not directly apply the paired-ads approach~\cite{Chen2019}.
For ad delivery, prior work shows that
  demographic parity for the delivery of a single ad
  cannot demonstrate lack of bias; instead,
  auditors must consider \emph{relative differences}
  in delivery of \emph{pairs} of ads~\cite{Ali2019a, Imana2021},
  as we describe in \autoref{sec:paired_ads_method}.
We expand on how our setting differs from prior studies on
  mitigating effect of inference error in \autoref{sec:related_work_mitigating_error}.

Our first contribution is to apply the use of inferred attributes
  to \emph{black-box audit of ad delivery algorithms} for skew
  that results in discrimination (\autoref{sec:infer_race_proofs}).
While the effect of attribute inference error has been studied in
  other domains~\cite{Zhu2022WeakPA, Elzayn2023, Ghazimatin2022},
  our work is the first to consider it for
  the domain of black-box
  auditing of ad delivery,
  where aggregate-only output and reliance on paired-ads requires new approaches.
We theoretically analyze and show how attribute inference error
  can lead to underestimating measurement of skew in ad delivery.
Our findings show that attribute inference error
  in the constructed audience,
  if not explicitly accounted for,
  can lead to failure to detect skew that exists in an ad delivery algorithm.
 
Second, we contribute an inference-aware approach to skew
  evaluation that
  corrects for inference error (\autoref{sec:new_skew_metric}).
We correct skew evaluation by modeling how
  inference error rates propagate from the targeted audience
  to the actual ad recipients.
Our model of inference error then allows us
  to estimate error in the delivery audience,
  and therefore in the evaluation of the platform's algorithms for potential skew.
Our approach can be generalized to any attribute inference
  method for which the inference error rates (defined in \autoref{sec:inf_error_def}) can be
  estimated on datasets with ground-truth demographics.
 
Our final contribution is 
  to demonstrate that our inference-aware
  evaluation is effective at detecting skew that we would
  otherwise miss when ignoring inference error (\autoref{sec:validation_summary}).
To validate our proposed correction,
  we estimate inference error on a real-world population
  and then use simulated ads to sweep through the space of parameters
  that reflect different possible conditions.
We show uncorrected data can fail to detect skew
  when the sample size available is small (\autoref{sec:vary_sample_size}) or
  when the true skew of the platform's algorithm is small but statistically significant (\autoref{sec:vary_threshold}),
   two conditions that are common in practice~\cite{Imana2021, Imana2024}.
We then apply our proposed solution to correctly
  detect skew under these conditions.
We use simulated ads to vary the true level of skew,
  a capability that is not feasible with real ads,
  and because exploring the full parameter space
  with real ads would be prohibitively expensive~\cite{Imana2021, Ali2019b}.

Our findings underscore that 
  attribute-inference methods are useful
  for detecting bias in ad delivery algorithms,
  but also that one should account for inference error
  when applying these methods to evaluate bias.
Our proposed method for inference-aware bias evaluation
  shows a path to expand auditing
  beyond a handful of U.S.~states whose voter records contain labeled demographic attributes~\cite{EACGovStateVoters}.
This advance relaxes the limitation of prior bias evaluation
  to these regions with demographic-rich voter datasets~\cite{Ali2019a, sapiezynski2022algorithms, Levi2022, Imana2024}.
It also provides a pathway to audit disparities across
  other protected attributes, such as gender or age,
  if inference probabilities can be estimated.
Moreover, our results suggests that the industry should
  carefully account for inference error when applying
  bias-correction methods to ad delivery, such as Facebook's VRS~\cite{Timmaraju2023}.

\section{Motivation and Problem Statement}
  \label{sec:problem_statement}

In this work, we propose
  applying use of inferred demographic attributes
  to \emph{paired-ads} methodology from prior work
  for auditing ad delivery algorithms.
In this setting an auditor only has external \emph{black-box access}
  to an ad platform (see \autoref{fig:decoupling}).
We focus on paired-ads methodology because it is the
  state-of-the-art method for auditing ad delivery
  that has been effective
  at uncovering algorithmic harms to individuals and society~\cite{Ali2019a, Ali2019b, Imana2021, Imana2024},
  supporting regulatory actions against platforms~\cite{FacebookvsHUD2},
  and pushing platforms to mitigate biases in their systems~\cite{FacebookCivilAuditProgress, Timmaraju2023}.
In this section,
  we motivate and expand on why we focus on adapting this specific methodology
  to use inferred demographic attributes.
 
\subsection{Need for Black-box Audits of Ad Delivery}
   \label{sec:need_for_external}

Black-box auditing has proved to be crucial
  for assessing harm
  in how ad delivery algorithms shape access to information and opportunities.
By ``black-box'' we refer to a setting where the auditor conducts an
  audit using only platform features available to any regular advertiser.
 
Prior black-box audits of ad delivery algorithms have uncovered
  biases and
  discrimination against protected demographic groups~\cite{Ali2019a, Imana2021, Imana2024, Ali2019b, Sankaranarayanan2023}. 
Starting with Sweeney's study in 2013~\cite{Sweeney2013},
  numerous studies hypothesized biased or discriminatory
  outcomes can be a result of platforms' algorithmic decisions,
  and not the targeting choices made by advertisers~\cite{Lambrecht2016, Dwork2019, datta2018discrimination}.
This hypothesis was proven by Ali and Sapiezynski et al.~by
  showing, via a black-box audit, that delivery of job and housing
  ads are biased by gender and race, even when an advertiser
  targets all demographic groups equally~\cite{Ali2019a}.
 This work served as a starting point for a DoJ lawsuit against Meta~\cite{FacebookvsHUDCase} and motivated a subsequent study that demonstrated the bias
   in job ad delivery can not be explained by differences in qualification
   of ad recipients~\cite{Imana2021}.
 Other follow-up studies showed
   the harm extends to other societally important domains such as
   politics~\cite{Ali2019b} and
   education~\cite{Imana2024}.
Given the far reaching effects of such harms,
  continued improvement of black-box auditing methods
  such as the one we propose in this work
  is an important step for keeping platforms accountable.

\subsection{Need for Paired-Ads Approach}
   \label{sec:paired_ads_method}

One technical challenge black-box auditors face is
  definitively attributing biased ad delivery to platforms' algorithms
  as opposed to market effects,
  differences in who is online, or other potential confounding
  factors.
For example,
 an ad may be delivered to less fraction of women than
 men because women are more expensive to reach~\cite{Lambrecht2016, Dwork2019}.
Using \emph{paired-ads} is a state-of-the-art methodology
  that has proven to be important for controlling for such factors~\cite{Ali2019a, Ali2019b, Imana2021, Imana2024},
  and therefore, is key to the uniqueness of the setting we study.
Conclusively demonstrating that platform-driven algorithmic decisions
  are the root cause of biased ad delivery is important for informing regulators
  tasked with enforcing anti-discrimination laws~\cite{datta2018discrimination}.  

In this methodology,
  an auditor runs a \emph{pair of ads}
  targeting the \emph{same audience} and \emph{at the same time}.
The auditor selects the paired ads based on some de-facto skew that the auditor
  hypothesizes the ad delivery algorithm will propagate.
For example, one may select ads for two jobs predominantly occupied by men and women,
  respectively, and hypothesize that a biased algorithm will show the ad for the predominantly men-occupied job
   to relatively more men, and vice-versa for the second ad.
The auditor tests for bias by comparing the \emph{relative difference} in how the
  two ads are delivered.
  
This setup, first proposed by Ali and Sapiezynski et al.~\cite{Ali2019a},
  is the only known approach to date to isolate the role of ad delivery algorithm
  for discrimination.
It controls for other confounding factors,
  such as market effects and differences in platform usage,
  by ensuring both ads are affected equally such that
  any relative difference between the two ads is attributable to choices made by the ad delivery algorithm.
Prior audits that did not rely on a paired-ads approach
  did not control for such relevant factors~\cite{Sweeney2013, Lambrecht2016}, and therefore,
  were not sufficient to be used by regulators
  to bring discrimination claims against platforms.
Platform-driven biases uncovered through the pared-ads
  approach ultimately led to
  the first successful legal action against Meta
  that led to the deployment of VRS~\cite{FacebookvsHUDCase,Timmaraju2023}.

One constraint to widely applying the paired-ads methodology is it requires knowing
  the demographic attributes of ad recipients,
  which may not be available to auditors.
To address this challenge,
  we explore the feasibility of using inferred attributes
  for conducting ad delivery audits using paired-ads approach.
We next discuss prior work related to the approach we study.

\section{Related Work}
	\label{sec:related_work}

Lack of access to demographic attributes, particularly those
  that can be deemed sensitive, such as race, gender, religion,
  disability status poses a challenge to auditing methods trying
  to assess algorithmic disparities based
  on protected characteristics~\cite{Andrus2021, Bogen2020, Holstein2019}. %
A report led by the Center for Democracy and Technology highlights the challenge remains despite increasing
  push by governments and policymakers to assess algorithmic
  systems for bias, and summarizes the various methodologies
  that practitioners currently use, such as collection and inference of attributes~\cite{Bogen2024}.
  
\subsection{Collecting Demographic Attributes}
  \label{sec:related_work_collecting}
In the context of black-box auditing ad delivery algorithms,
  collecting demographics from voter datasets is a
  commonly used approach~\cite{Speicher2018, Ali2019a, Imana2021, Levi2022, Imana2024}.
Other options, such as collecting data from volunteers~\cite{MozillaRally}, have been tried but have not gained as much traction as the voter datasets approach (the latter
 are available at a lower cost and
  are easily accessible).
However, there are only a few states in the U.S.~whose publicly available
  voter datasets contain demographic attributes such as race~\cite{EACGovStateVoters}.

Platforms conducting internal audits of their algorithmic systems can
  request users to voluntarily self-identify
  their demographic attributes to support fairness efforts,
  as seen with Meta and LinkedIn~\cite{metatech, LinkedInSelfID}.
However, many platforms choose not to collect demographic
  information at account creation due to privacy concerns
  and the potential misuse of data~\cite{metatech}.
Other platforms such as Apple have used federated approaches
  to analyze demographic data on users' devices to avoid
  centrally collecting data~\cite{Villenueve2023}.
Our work focuses on external black-box auditing
  without access to self-identified demographic data
  collected by platforms.

\subsection{Inferring Demographic Attributes}
  \label{sec:related_work_inferring}
Another common approach is to \emph{infer} demographics
  when external data sources or self-identification is not possible.
BISG is a commonly used method for inferring
  race from name and location~\cite{Elliott2009},
  and has been applied to examine disparities in
  lending~\cite{CfpgBisg2014, Baines2014},
  healthcare~\cite{Adjaye2014},
  tax auditing~\cite{Elzayn2023},
  mortgage pricing~\cite{Zhang2018}, and
  human-mobility~\cite{Luo2016}.
Other studies have proposed
  transfer learning from domains for which demographic
  data is already available~\cite{Ashurst2023, Islam2024},
  or using machine-learning to infer attributes~\cite{Mccartan2024, Chen2024}.
Another method infers gender from first name~\cite{liu2013s}.
Airbnb uses human labeling to infer perceived race from names
  and photos of users, 
  and uses inferred race to
  measure racial discrimination against its users~\cite{Basu2022}.
LinkedIn infers user age from graduation dates,
  and gender from first names,
  but they do not disclose details of their mechanism~\cite{LinkdeinInfer}.
We too infer demographics,
  but are the first to explore its use for
  black-box audit of ad delivery.
  
More closer to our work is the use of BISG to evaluate racial disparity in Meta's
   VRS~\cite{metatech}.
Following a historic settlement between Meta
  and the U.S. Department of Justice~\cite{FacebookvsHUD2},
  Meta agreed to implement VRS
  to address unfairness in the delivery ads for economic opporunities~\cite{Timmaraju2023}.
The system was originally deployed for housing
  ads in January 2023,
  and later that year for employment and credit ads~\cite{FacebookvsHUD2023}.    
Meta employs BISG to estimate the racial composition
  of the delivery audience of an ad and make adjustments
  accordingly.
While Meta's publication acknowledges
  BISG's misclassification rate,
  it is not clear if their VRS implementation considers such error,
  and their paper does not explicitly consider it.
Unlike Meta's internal evaluation,
  we explore how to factor inference error
  in external black-box auditing of ad delivery,
  providing a method to account for inference error.
  
 \textbf{Effects of Ignoring Inference Error:} 
Although tools to infer demographics such as race and gender
  are widely used,
  they are known to have high misclassification rates~\cite{Ashurst2023, Lockhart2023}.
Prior studies have shown that uncertainty in demographic attributes
  may lead to inaccuracy in both building fair algorithms~\cite{Shah2023, Ni2024}
  and measuring demographic disparities~\cite{Baines2014, Chen2019}.
Another study used inferred race in fair-ranking
  algorithms,
  and showed inference produces
  unfair rankings by skewing the demographics
  represented in the top-ranked results unless the race inferences
  are highly accurate~\cite{Ghosh2021}.
Rieke et al.~also show that race inference methods lead to
  significant error in the magnitude of estimates of racial disparities
  among Uber users,
  either underestimating or overestimating these disparities~\cite{Rieke2022}.
However, they demonstrate the methods may still be useful in detecting  the direction of disparity.
In our work,
  we also show ignoring inference error can lead to
  underestimating skew in ad delivery.
We additionally propose a method to correct for inference error
  for the paired-ads approach.

\subsection{Mitigating Effect of Error in Demographic Attributes}
  \label{sec:related_work_mitigating_error}
A number of studies have proposed methods to correct for
  inference error when estimating algorithmic disparities,
  but they do not apply to our restricted setting of black-box
  auditing ad delivery using paired-ads.
Our setting has two unique constraints. First, only aggregate data is
  available during model evaluation, as illustrated in \autoref{fig:decoupling}.
The auditor does not have access to
  the individual-level uncertainty of attributes that most approaches rely on
  for model evaluation.
Second, because our method relies on evaluating relative differences
  between a pair of ad campaigns,
  one cannot apply standard group fairness metrics on a single ad to evaluate bias.
In a line of work that studies the effect of inference error
  on algorithmic audits, 
Chen et al.~derives the statistical bias in estimating algorithmic
  disparity using inferred demographic attributes for one
  group fairness metric: demographic disparity~\cite{Chen2019}.
Wastvedt et al.~generalizes their approach to extend to other popular
  group fairness notions~\cite{Wastvedt2024}.
The statistical bias they demonstrate motivates the need to
  account for inference error in our work.
However, their specific analysis based on group fairness metrics does not apply
  to our specific setting that evaluates the relative
  difference in delivery between two ads.
One cannot test a group fairness metric on a single ad
  to evaluate bias in ad delivery due to confounding factors
  (see \autoref{sec:paired_ads_method}).

Another group of work proposes methods to correct for inference
  error when evaluating algorithmic biases.
Concurrent to our work, two working papers propose a method to estimate
  disparities while adjusting for noise in inferred demographic attributes~\cite{Lu2024, Mccartan2024}.
While the goal of these studies is similar to ours,
  they use an approach that relies on propagating
  the uncertainties in demographic attributes
  to model evaluation.
Elzayn et al.~also use probabilities of raw inference
  to bound the true racial disparity
  given estimates based on inferred attributes~\cite{Elzayn2023, Elzayn2023b}.
Ghazimatin et al.~identifies how true
  algorithmic disparity can be estimated using inferred attributes,
  but unlike our setting, they focus on fairness in ranking~\cite{Ghazimatin2022}.
Zhu et al.~propose using a family
   of demographic inference methods to debias the estimate of true algorithmic
   disparity~\cite{Zhu2022WeakPA}.
Compared to these studies,
  our work addresses the stricter 
  requirements of black-box auditing,
  where only aggregate data is
  available to the auditor during model evaluation.
This requirement makes it infeasible to 
  apply corrections that track specific individuals, their inferred attributes,
  and corresponding probabilities.

Another line of work studies how to build fair algorithms
  while accounting for noise in demographic attributes.
While these studies consider similar models of noise
  in demographic attributes, they address a
  different problem of \emph{training} a fair algorithm instead of
  \emph{auditing} an existing algorithm.
Fair-classification is one prominent domain where noise-tolerant
  training algorithms have been proposed~\cite{Ghosh2023}.
Beyond fair-classification, other studies have explored the problem of
  noisy demographic attributes for fair subset-selection~\cite{Mehrotra2021} and fair-ranking~\cite{Ghosh2021, Mehrotra2022}.  
Lamy et al.~proposes an approach for fair-classification
  with noisy binary demographic attributes that works by adjusting the desired
  ``fairness tolerance'' based on estimates of noise in the attributes~\cite{Lamy2019}.
Celis et al.~extends noise-tolerant fair-classification
   to non-binary noisy demographic attributes~\cite{Celis2021}.
A related study by Awasthi et al.~identifies specific conditions for
  noise in demographic attributes under which
  a classifier's fairness can be ensured~\cite{Awasthi2019}.
These studies on noise-tolerant fair-classification
   are similar to our work in that they consider the effect
   of noisy demographic attributes
   and consider a group-level noise model
  where the error in the attributes is the same for all individuals
  in a single (inferred) demographic group.
But they consider group fairness metrics that cannot be directly
  applied to black-box auditing ad delivery where disparity is
  measured using aggregate data and
  by looking at relative performance of a pair of ads.

\section{Adapting Paired-Ad Auditing to use Inferred Attributes}
  \label{sec:setup_inf_race}

Our approach combines paired-ad auditing
  with inferred demographics.
We cannot then simply infer demographic attributes and use the auditing result as-is,
  since inference methods are known to have error.
We therefore next build a model of inference error
  and how that error propagates through to skew evaluation.

\subsection{Setup and Notations: Auditing with Paired Ads}
  \label{sec:setup_notations}

\reviewfix{}
We use a black-box auditing setting from prior work where where the auditor
  runs a \emph{pair} of ads relying on features available to any regular advertiser~\cite{Ali2019a, Imana2021}.

The auditor runs both ads targeting the same audience.
Let set $U$ represent the audience.
We consider an audience composed of two demographic
  groups, A and B.
We assume group A is a ``disadvantaged''
  group for which harm in terms of over- or under-exposure of ads
  must be minimized and
  group B is the ``advantaged'' group.
These groups can be defined based on a demographic
  attribute of interest to the auditor, such as race, gender, or age.
For simplicity, we focus on two primary groups,
  denoted as Group A and Group B, and
  categorize all remaining users as ``Other.''
We use the subscripts ``a,'' ``b,'' and ``o'' to represent
  users belonging to Group A, Group B, and the
  Other category, respectively. Extending the analysis to
  more than two groups is left for future work.

Let $u_{a,*} $ and  $u_{b,*}$
  represent the number of individuals in the targeted audience
  that are in group A and B, respectively.
We use ``*'' as a placeholder for a subscript that indicates
 whether the true demographic attribute is available and used by
 the auditor (subscript $t$),
 or is inferred from other available information (subscript $i$).  
The target audience $U$ typically contains equal number of individuals
  from each demographic group, following standard practice
    in prior ad delivery audits~\cite{Ali2019a, Imana2021}.
 Therefore, the following holds: $u_{a,*} = u_{b,*} = \frac{|U|}{2}$.

After the ads start running, the auditor collects demographic
  breakdown of ad impressions.
For attributes that platforms do not report,
  auditors rely on proxies~\cite{Ali2019a, Imana2021},
  a detail we omit in the rest of this work.
Let $n_{1,a,*} $ and  $n_{1,b,*}$
  represent the number users that saw the first ad and are in group A
  and B, respectively.
$n_{2,a,*} $ and  $n_{2,b,*}$ are similarly defined for the second ad.
Again, the subscript ``*'' is placeholder for whether the auditor
  is working with inferred or true demographic attributes.
The auditor then calculates the fraction of users in group A
that saw each ad as:
$ s_{1,a,*} =$ $\frac{{n_{1,a,*}}}{ n_{1,a,*} + n_{1,b,*}}$  and $s_{2,a,*} =$ $\frac{{n_{2,a,*}}}{n_{2,a,*} + n_{2,b,*}}$.
In the absence of a skewed ad delivery algorithm,
  the auditor expects: $s_{1,a,*} = s_{2,a,*}$.
Let $D_*$ represent the skew in ad delivery,
  which is given by:
 $$D_* = s_{2,a,*} - s_{1,a,*}$$
 
To testing statistical significance of the skew,
  the auditor applies a Z-test for difference in
  proportions to test whether there is statistically significant difference
  between the fraction of individuals in group A that saw each ad ($s_{1,a,*}$ and $s_{2,a,*}$).
The null hypothesis is that $D_*=0,$ and the alternate hypothesis is $D_*>0$.
The test-statistic is given by:
\begin{equation}\label{eq:generic_Z_formula}
\begin{aligned}
& \myV{ZTEST}(n_{1,a,*}, n_{1,b,*}, n_{2,a,*}, n_{2,b,*}) = \frac{D_{*}}{{ {\myV{SE}_* }}}
  \end{aligned}
\end{equation}
where ${\myV{SE}_*} = \sqrt{\hat{s}_{a,*} (1 - \hat{s}_{a,*}) \left(\frac{1}{n_{1,*}} + \frac{1}{n_{2,*}}\right)}$,
$n_{1,*} = n_{1,a,*} + n_{1,b,*}$ and $n_{2,*} = n_{2,a,*} + n_{2,b,*}$ and
$\hat{s}_{a,*}$ is the fraction of users in group A, either true or inferred,
 in combined set of all people that saw at least one of the two ads:
 $\hat{s}_{a,*} = \frac{n_{1,a,*} + n_{2,a,*}}{n_{1,*} + n_{2,*}}$.
At a level of significance $\alpha$ (typically, $0.05$) and a corresponding critical value
  of $Z_\alpha$ from the Z-table
  for standard normal distribution,
  the auditor concludes that there is a statistically significant skew
  in the ad delivery algorithm if the test statistic is greater than $Z_\alpha$.
We summarize these notations in \autoref{tab:delivery_notation_with_correction}.


\subsection{Inference Error in Audience Construction} 
  \label{sec:inf_error_def}

\begin{figure}
  \centering
   \includegraphics[width=0.99\columnwidth]{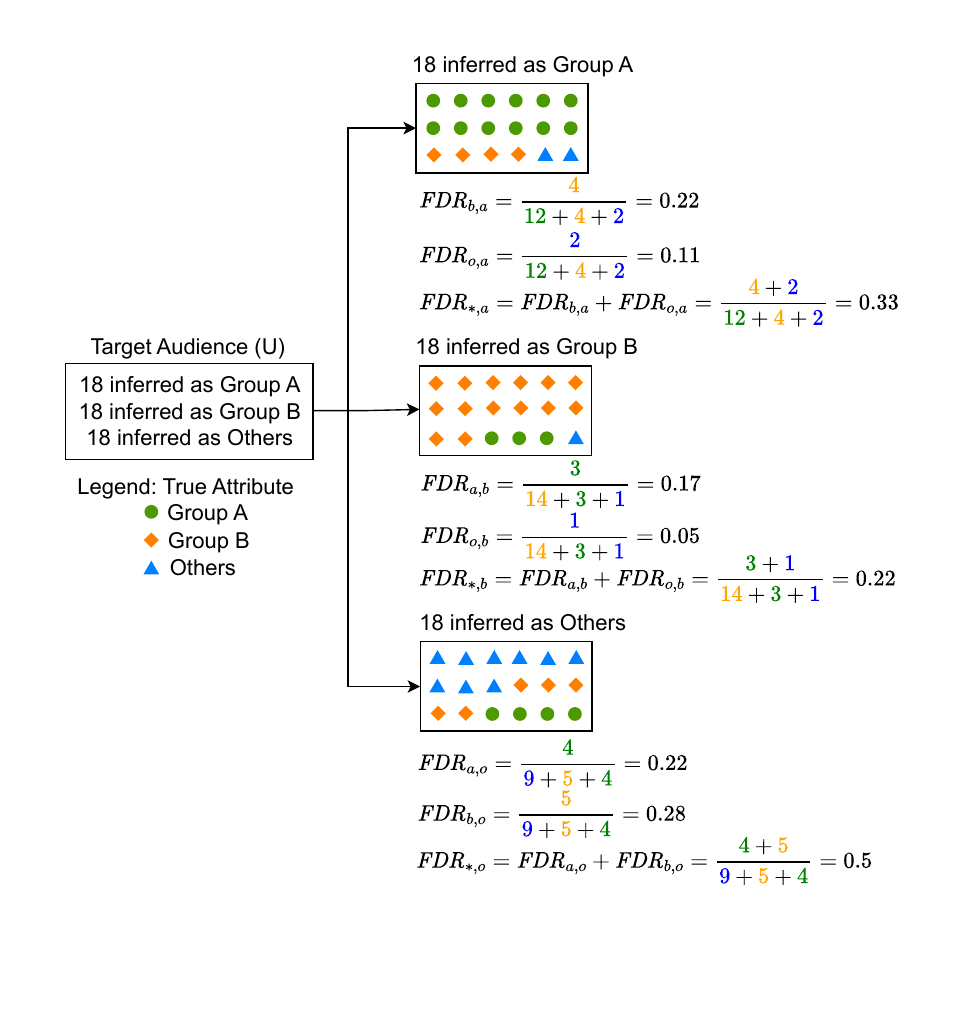}
    \caption{An illustration of how False Discovery Rates are calculated for an audience constructed with inferred race. All values shown are fractions of individuals.}
      \label{fig:bisg_error_illustration}
\end{figure}

We next consider inference error in an audience $U$ constructed
  by an auditor using inferred demographic attributes.
First,
  some individuals inferred as group A might not actually
  be in group A.
The upper box of \autoref{fig:bisg_error_illustration} illustrates this error.
The box represents \emph{population inferred as group A},
  but some are of other groups---here it is the bottom row of orange diamond and blue triangles.
We define the \emph{False Discovery Rate 
  for individuals in group A}, or $\myV{FDR}_{*,a}$,
  as the ratio individuals that are not in group A but are inferred as group A.
Since we consider only two demographic groups (group A'', ``group B''),
  and group all others as ``Other'',
  this error is a sum of the ratio of
  individuals in group B ($\myV{FDR}_{b,a}$; orange diamonds)
  and ``Other'' individuals ($\myV{FDR}_{o,a}$; blue triangles)
  among the population inferred as group A.  
We analogously define False Discovery Rates for
  individuals in group B ($\myV{FDR}_{*,b}$) and in ``Other'' ($\myV{FDR}_{*,o}$)
  as illustrated in the middle and bottom box of the figure,
  respectively.
The relationship between the different rates is given by:
  \begin{equation}\label{eq:error_FDR_def}
  \begin{aligned}
  &\myV{FDR}_{*,a} = \myV{FDR}_{b,a} + \myV{FDR}_{o,a} \\
  &\myV{FDR}_{*,b} = \myV{FDR}_{a,b} + \myV{FDR}_{o,b} \\
  &\myV{FDR}_{*,o} = \myV{FDR}_{a,o} + \myV{FDR}_{b,o}
  \end{aligned}
\end{equation}

We model inference error by assuming it is
  is exactly some expected rate that is estimated
  from an auxiliary dataset with ground truth race.
We give details on how we estimate these error
  rates in practice in \autoref{sec:validation_summary}.

\subsection{Assumptions on Skew In Ad Delivery}
 \label{sec:setup_for_though_exp}
 
To evaluate the effect of inference error on skew evaluation,
    we use the following parameters and assumptions about
    skew in ad delivery (see below for justification).
    
\reviewfix{}  
\textbf{Base delivery rate ($R$):} We assume ads are delivered to a fixed fraction, $R$, of the targeted audience $U$.
This assumption reflects legal requirements for
  consequential domains, such as those related to employment
  and housing, which mandate that individuals have a fair chance
  of seeing an ad regardless of their true or inferred demographic
  attributes.
Thus, for any individual in a targeted audience that is a potential ad recipient,
  $R$ is the probability they receive the ad.
In our setup, $R$ operates at the individual level, representing the
  probability that any given person within the audience sees the ad
  under fair delivery.
 We capture any deviations from this uniform delivery probability using
   a skew parameter.

\textbf{Skew parameter ($S$):} 
We add a parameter $S$ to model any skew introduced by the delivery algorithm;
We assume the algorithm makes decisions on the basis of
  true demographic attributes.

Among the two groups A and B,
  the algorithmic skew $S$ will alter the base delivery rate $R$
  in favor of one of the demographic groups,
  the ``advantaged'' group (group B), over
  the ''disadvantaged'' group (group A).
Our simplifying assumption of focusing on two groups requires
  assigning individuals incorrectly inferred to be in
  group A or B, but who actually belong to ``Other'', to
 either the advantaged or disadvantaged group.
However, our setup is flexible and allows this assignment
  to be made based on the specific demographic attribute
  and problem domain.

The skew parameter $S$ captures whether ad delivery is biased
  in favor of or against a disadvantaged group (group A).
  $S=1$ means the algorithm has no skew,
  $S<1$ means it delivers fewer ads to the disadvantaged group,
  $S>1$ means means it over-delivers to the disadvantaged group.
When the ad being advertised is for an economic
  opportunity, such as a job for which
  equal access is important,
  we consider the algorithm as discriminatory if delivery is biased against the disadvantaged group ($S<1$).
If the ad is considered harmful to users,
  in which case we consider delivery that is biased towards a disadvantaged group ($S > 1$) to
  be discriminatory.
In this work, we focus on the former scenario where the ad is
  for an economic opportunity.
  
We expect the two ads to show different skews,
  and so model the first ad as skewed against the disadvantaged group ($S < 1$) and the second as not ($S=1$).
This model reflects a first ad
  that reflects a demographic bias that is captured by the platform's algorithms,
  causing skewed delivery.
We assume the ad creative chosen for the second ad is neutral.
Prior work shows a real-world example
  that reflects this assumption,
  where an ad for Hip-Hop music (expected to be skew Black)
  and a general ad for music (expected to be neutral) are compared~\cite{Ali2019a}.  
  
 Formally, $S$ is a multiplier representing increased
   or decreased delivery rate of the first ad:
   $R \times S$ for individuals in the disadvantaged group A,
   and $R \times (2-S)$ for individuals in the advantaged group B.
 We assume $0 < S < 2$, where $S=1$ represents a case where
   there is no skew in ad delivery algorithm.
 For example, if $R=0.1$ and $S = 0.87$,
   then the delivery rate for the disadvantaged group
   for the first ad will be $0.87 \times 0.1 = 0.087$ and
   the rate for the advantaged group for the first ad will be
   $1.13 \times 0.1 = 0.113$. The rate for both groups does
   not change for the second ad and remains at 0.1.

Our assumption that $S$ operates on true attributes rather than inferred attributes
  is partially supported by Meta's current implementation of VRS for reducing racial bias
  in ad delivery.
As part of the VRS settlement requirements,
  Meta reports that they use BISG-inferred race for measuring skew
  in delivery.
However, their ad delivery decisions rely on user embeddings
  built from Meta's comprehensive data, not BISG
  (see Section 3.3 in~\cite{Timmaraju2023}).
Because these embeddings reflect rich behavioral signals
  correlated with race, the algorithm's behavior is likely
  closer to conditioning on true race than on BISG-inferred race.
These practices support our simplifying assumption that $S$
  captures algorithmic skew relative to true demographic attributes.
\reviewfix{}
Therefore, we do not model any dependence between $S$
  and inferred attributes, and leave a joint treatment of delivery skew
  and demographic inference uncertainty to future work.


\section{Effects of Attribute Estimation on Auditing Ad Delivery}
  \label{sec:infer_race_proofs}
  
We next show that using inferred attributes
  to evaluate ad delivery algorithms
  can underestimate the true level of skew and
  thus miss detection of algorithm-induced skew.

We show how inference error can affect evaluation of skew in ad delivery
  through two theorems.
In \autoref{thm:error_no_effect},
  we consider a baseline case where there is no algorithmic skew
  and, as a result, inference error does not affect evaluation of skew.
In \autoref{thm:error_underestimate_skew},
  we consider a case with known amount of skew.
We show that measuring skew using inferred attributes
  underestimates the algorithm's true skew,
  had it been
  measured using true demographic attributes.
These results apply for any
  attribute inference method with known error rates ($\myV{FDR}_{*,*}$ defined in \autoref{sec:inf_error_def}),
  and assume the specific behavior of skew described in \autoref{sec:setup_for_though_exp}.
Together, these results support our claim that attribute inference
  must be used carefully with consideration
  for how inference error affects evaluation of
  skew in ad delivery algorithms.

\begin{table}
\begin{center}
\begin{tabular}{rll|c}
\hline
Number of & Group A targeted & & $u_{a,*}$  \\
 & Group B & & $u_{b,*}$ \\
 & Others & & $u_{o,*}$ \\
\hline
\multicolumn{2}{l}{Total number of people seeing} &  ad 1 & $n_{1,*}$  \\
 & & ad 2 & $n_{2,*}$ \\
Number of & Group A seeing & ad 1 & $n_{1,a,*}$  \\
 & Group B & ad 1 & $n_{1,b,*}$  \\
 & Others & ad 1 & $n_{1,o,*}$  \\
 & Group A & ad 2 & $n_{2,a,*}$ \\
 & Group B & ad 2 & $n_{2,b,*}$  \\
 & Others & ad 2 & $n_{2,o,*}$  \\
\hline
Fraction of & Group A who saw & ad 1 & $s_{1,a,*}$    \\
 & Group A & ad 2 & $s_{2,a,*}$  \\
\hline
\multicolumn{3}{l}{Skew between ads} & $D_{*}$  \\
\multicolumn{3}{l}{Test statistics} & $Z_{*}$  \\
\end{tabular}  
\end{center}
\caption[Notations for measurement of skew in ad delivery.]{\reviewfix{}Notations for measurement of skew in ad delivery.
The ``*'' in each is a placeholder for an audience with:
$t$: true race;
$i$: inferred race, ignoring inference error;
$c$: inferred race with omniscient correction;
$f$: inferred race with our solution to accounts for expected inference error.}
  \label{tab:delivery_notation_with_correction}
\end{table}

\subsection{First Case: No Algorithmic Skew}

We first consider a simple case where there is no algorithmic skew.
In this case,
  we show that measuring skew using inferred demographic attributes does
  not affect our conclusion of skew:

\reviewfix{} 
\begin{theorem}\label{thm:error_no_effect}
If an ad delivery algorithm is not skewed by a protected demographic attribute ($S = 1$), inference error does not affect the measurement of skew in ad delivery.
Specifically, the skew an auditor measures is $0$ in both cases
 where the auditor targets using true attributes ($D_t = 0$) and inferred attributes ($D_i = 0$).
\end{theorem}
The theorem shows that when there is no ad delivery skew,
  measurement of skew ($D_i = 0$) is correct regardless of inference error.
We prove the theorem in
  \autoref{sec:appendix_proof_inf_error_effect}, and, in \autoref{sec:exp_1_sec},
  illustrate this case in a thought experiment.
We find that inference error affects both ads equally,
  and because we measure skew as \emph{relative} difference in delivery between the ads,
  we see no overall change in our conclusion.

\subsection{Second Case: Skew is Underestimated}
  \label{sec:second_case_underestimate}
In a second case,
  we add a known amount of algorithmic skew and analyze
  how it affects measurement of skew in ad delivery.
We show that, if the ad delivery algorithm is skewed,
  measuring skew using inferred attributes underestimates
  the true skew in the algorithm.

\begin{theorem}\label{thm:error_underestimate_skew}
If an ad delivery algorithm is skewed by a protected demographic attribute ($S \neq 1$),
  the skew that an auditor measures by targeting using inferred attributes ($D_i$) underestimates
  the true skew one would measure using true attributes ($D_t$): $|D_i| < |D_t|$.
\end{theorem}

We prove the result in \autoref{sec:appendix_proof_inf_error_effect}, and in \autoref{sec:exp_1_sec},
  we provide a concrete example where underestimation
  hides a skew that exists by making it appear as statistically insignificant.
The intuition behind this result is that inference error 
  always pushes towards a neutral outcome,
  reducing how much skew is observed.
This intuition follows from our test process with paired ads:
  skew is maximized by targeting audiences using true demographic attributes.
Targeting with inferred attributes produces an actual audience composed of a mix of
  the true attributes.
Because we assume algorithm skew operates on true attributes,
  the effect of the algorithmic skew is reduced because it only
  applies to a subset of the mixed audience that matches the
  true demographic attribute.
This claim, that error only underestimate skew, is a contrary to our initial
  assumption that error could both hide or exaggerate skew.
We developed this intuition from examples in \autoref{sec:infer_race}.

This result demonstrates that 
  auditing ad delivery algorithms
  for bias using inferred attributes 
  can underestimate true level of racial bias, $|D_i| < |D_t|$.
If demographics are inferred, one must consider how error may change 
  their use.
This result also suggests that use of attribute inference
  in real-world applications, such as Meta's VRS~\cite{Timmaraju2023}
  should be examined closely.
Because our methodology is quite different than their application
  (we use paired ads and they do not, for example),
  further examinations are future work.


\section{Inference-aware Auditing for Skew}
  \label{sec:new_skew_metric}

Having shown that inference error can hide skew during auditing (\autoref{sec:infer_race_proofs}),
  we next suggest how to account for such error. 
The challenge is that it is difficult to track
  how error propagates
  from the targeted audience to the delivery audience
  because the fairness evaluation is decoupled
  from the target audience (\autoref{fig:decoupling}).
Detecting such propagation is hard because platforms do not
  provide demographics
  of specific  ad recipients, only aggregate statistics.
A second challenge with modeling how inference error
  propagates to the delivery audience is that the parameters
  we use to model algorithmic skew ($S$)
  and the delivery rate ($R$)
  are not known in practice.

Our insight for inference-aware skew evaluation,
  even with limited ad delivery statistics,
  is that we can solve for $R$ and $S$ and
  model how inference error propagates
  based on the aggregate data that we get
  from the platform.
We can then adjust our detection sensitivity accordingly.

\subsection{Modeling Propagation of Inference Error}
  \label{sec:error_propagation_model}

In order to model how inference error propagates to ad delivery,
  we first solve for $R$ and $S$.
The following theorem gives a closed-form solution for both
  parameters (we prove the theorem in \autoref{sec:solve_R_and_S_app}).
 
 \reviewfix{}
\begin{theorem}\label{thm:solve_for_S}
Assuming we can estimate FDRs of the attribute-inference method based on a dataset with ground truth, the targeted audience $U$ is constructed so that it contains an equal number of individuals inferred as group A and inferred as group B, and assuming the specific behavior of skew (described in \autoref{sec:setup_for_though_exp}),
we can solve for the delivery rate ($R$) and the skew parameter ($S$)
  as follows:
$R = \frac{XP - MY}{NP-MQ}$ and $S = \frac{XQ-NY}{MY-XP}$,
where
$M = \frac{|U|}{2} - (|U| \times \myV{FDR}_{*,a})$, 
$N = |U| \times \myV{FDR}_{*,a}$,
$X =  n_{1,a,i}$, 
$P = |U| \times \myV{FDR}_{a,b} - \frac{|U|}{2}$,
$Q = |U| - |U| \times \myV{FDR}_{a,b}$, and
$Y = n_{1,b,i}$.

\end{theorem}

Using $R$ and $S$,
  we can model how inference error propagates from the targeted
  audience to the delivery audience.
As an example, we describe below how the error rate propagates
  to the delivery audience of the first ad.

Among those inferred as group A in the targeted audience ($u_{a,i}$),
  we expect $u_{a,i} \times \myV{FDR}_{b,a}$ people
  to actually be in group B.
To know how many of those group B individuals see the first ad,
  we multiply their expected number ($u_{a,i} \times \myV{FDR}_{b,a}$) by the delivery rate ($R$)\
  and the skew applicable for individuals in group B ($2-S$).
We then divide the resulting number by the total number
  of individuals inferred as group A that saw ad 1 ($n_{1,a,i}$) to derive
  the ratio of individuals that are in group B,
  which we denote by $\myV{fdr}_{b,a,1}$.
Compared to the notation for targeted audience,
  we use lower case ``$\myV{fdr}$'' to represent error in the delivery audience
  and an additional subscript (1 or 2) to indicate which ad:
\begin{equation}\label{eq:bisg_out_error}
\begin{aligned}
\myV{fdr}_{b,a,1} &= \frac{u_{a,i} \times \myV{FDR}_{b,a} \times R \times (2-S)}{n_{1,a,i}}, 
\end{aligned}
\end{equation}

We can apply a similar procedure to derive the other
  error rates $\myV{fdr}_{o,a,1}$, $\myV{fdr}_{a,b,1}$, and $\myV{fdr}_{o,b,1}$.
Similar to the error in the targeted audience (\autoref{eq:error_FDR_def}),
we define the following notations that represent the total
  ratio of ad recipients wrongly labeled as being in group A
  ($\myV{fdr}_{*,a,1}$) or in group B ($\myV{fdr}_{*,b,1}$):
$\myV{fdr}_{*,a,1} =\myV{fdr}_{b,a,1} + \myV{fdr}_{o,a,1}$ and  $\myV{fdr}_{*,b,1} =\myV{fdr}_{a,b,1} + \myV{fdr}_{o,b,1}$.

\subsection{Correction Based on Expected Error}
  \label{sec:our_correction}
We next adjust ad delivery statistics using
  our model of how error propagates.
Among those that saw an ad,
  our goal is to derive an estimate of the
  number of individuals in true group A ($n_{1,a,f}$) and in true group B ($n_{1,b,f}$)
  based on the number of individuals in inferred group A ($n_{1,a,i}$)
  and inferred group B ($n_{1,b,i}$).
  
As summarized in \autoref{tab:delivery_notation_with_correction},
  we use ``$f$'' to denote the corrected
  statistics we compute
  based on the expected inference error.
Using our model of how error propagates to ad delivery audience (\autoref{sec:error_propagation_model}),
 we calculate the corrected ad delivery statistics
  as follows:

\begin{equation}\label{eq:corrected_stats}
\begin{aligned}
n_{1,a,f} = n_{1,a,i} \times (1 - \myV{fdr}_{*,a,1}) + n_{1,b,i} \times \myV{fdr}_{a,b,1} \\
n_{1,b,f} =  n_{1,a,i} \times \myV{fdr}_{b,a,1} + n_{1,b,i} \times (1 - \myV{fdr}_{*,b,1})  \\
n_{2,a,f} = n_{2,a,i} \times (1 - \myV{fdr}_{*,a,2}) + n_{2,b,i} \times \myV{fdr}_{a,b,2} \\
n_{2,b,f} =  n_{2,a,i} \times \myV{fdr}_{b,a,2} + n_{2,b,i} \times (1 - \myV{fdr}_{*,b,2})  \\
\end{aligned}
\end{equation}
  
We then apply a hypothesis test for significance
  of skew in ad delivery using
  our corrected delivery statistics.
We plug these values to \autoref{eq:generic_Z_formula},
  to calculate the test statistic for racial skew that
  accounts for inference error: $Z_f = \myV{ZTEST}(n_{1,a,f}, n_{1,b,f}, n_{2,a,f}, n_{2,b,f})$.

\section{Validation of Proposed Correction}
  \label{sec:validation_summary}

Finally, we validate our proposed method for inference-aware skew evaluation.
We simulate various levels of skew in an ad delivery algorithm
  and compare the outcome of evaluation with and without
  correcting for error.
We find that when the skew in the ad delivery algorithm
  and the sample size available for auditing are both small,
  inference error hides skew from the auditor
  as statistically insignificant.
In contrast,
  our inference-aware skew evaluation corrects the expected size of each group in 
  the delivery audience we obtain, which reflects on the statistical tests
  on the delivery audience, 
  making detection possible even if skew is small.

We estimate inference error rates on a real-world population and
  use simulated ads to sweep the parameter space and consider
  how our proposed correction affects the evaluation of skew.
We use simulations for two reasons.
First, in simulations we know ground truth
  and so we can make strong statements on how inference error affects outcomes,
  while ground truth is unknown or difficult to obtain in real-world experiments.
Second, simulation allows us to sweep the parameters space
  to understand how sensitive our results are to many possible conditions.
Such wide exploration would be expensive with paid, real-world advertisements,
  as shown in prior work~\cite{Imana2021, Ali2019b}.

\subsection{Validation Methodology}
  \label{sec:validation_setup}

  \begin{figure}
  \centering
   \includegraphics[width=0.95\columnwidth]{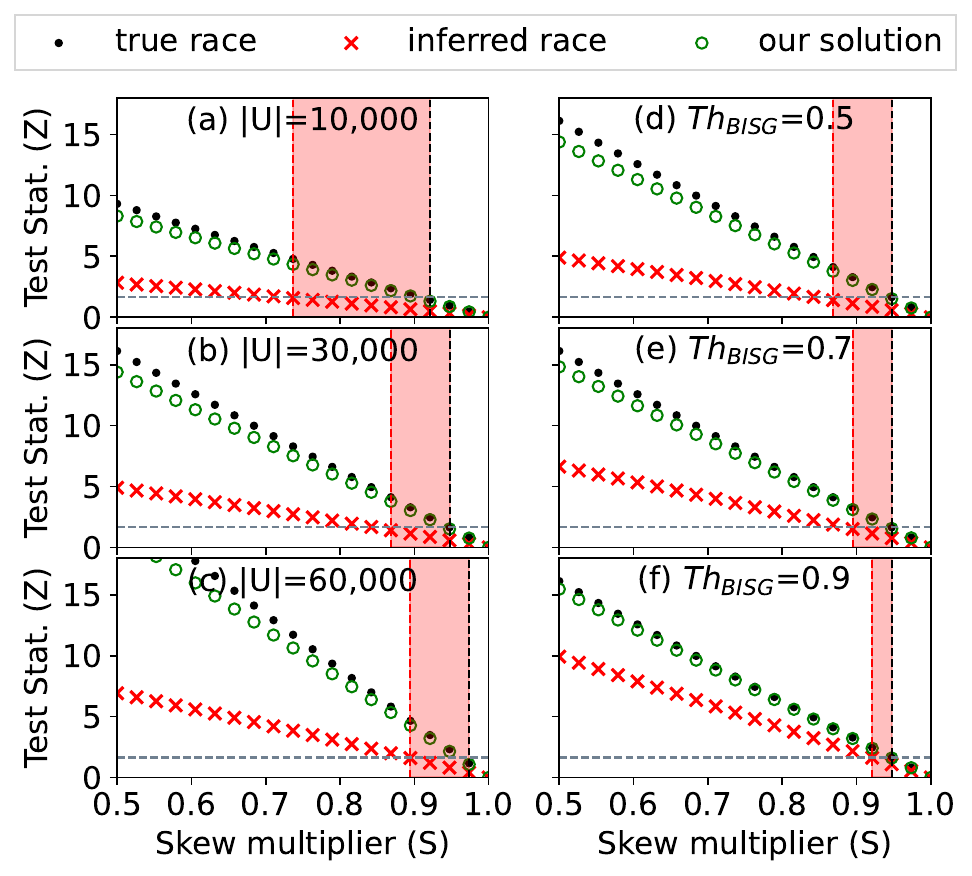}
    \caption{Left column shows the effect of sample size on our inference-aware skew evaluation.  Parameters: $R=0.065$, $\myV{Th}_{\Vs{BISG}} = 0.5$, $\myV{FDR}_{b,a}=0.4727,$ $\myV{FDR}_{o,a}=0.030$ $\myV{FDR}_{a,b}=0.144$, $\myV{FDR}_{o,b}=0.032$. The right column shows the effect of BISG inference error rates on our inference-aware evaluation.  Parameters: $R=0.065$, $|U|=30,000$. In both columns, as sample size and BISG threshold increases (inference error decreases), the red shaded region where inference error leads to hiding skew that exists gets reduced.}
      \label{fig:vary_all_figures_combined}
\end{figure}

Our validation methodology uses simulations
  where we sweep the amount of skew ($S$)
  through a wide range of possible values,
  then consider how detectable the skew is through
  three different statistics (\autoref{eq:generic_Z_formula}).  
The first statistic represents skew measured by targeting
  with true demographic attributes,
  and serves as ground truth for presence of skew
  in the platform's ad delivery algorithm.
The other two statistics are measured by building the target audience
  with inferred attributes and after adjusting for inference error (\autoref{sec:our_correction}).
  
For concreteness in our validation experiments,
  we consider race as our attribute of interest.
We use the group A and B defined in \autoref{sec:setup_notations}
  to represent ``Black'' and ``White'' racial groups, respectively.
We use BISG for inferring race~\cite{Elliott2009}.
We instantiate the inference error rates defined in \autoref{sec:inf_error_def} by applying BISG
  to real-world voter datasets that contain name and location.
While we use race and BISG to validate our methodology,
  one can apply our approach
  for other inference methods for which these
  error rates can be estimated.

For BISG, we assign a race when the inference probability
   exceeds a threshold.
We denote this threshold by $\myV{Th}_{\Vs{BISG}}$.
Another common approach is to directly use the raw probabilities~\cite{Elliott2008, Chen2019},
  but we use thresholding because
  we must construct audiences that match specific racial demographics.
In addition,
  since the platform only provides \emph{aggregate results}
  regarding to whom an ad was shown,
  we cannot propagate the raw probabilities through to our evaluation of potential skew.

The rate of inference error depends not only on the BISG method,
  but also on the audience it is measured on.
It is important to measure the error rate with respect to
  the same audience that will be used in a real-world experiment
  because that is the error relevant for the experiment.
Here we measure the inference error using the North Carolina dataset,
  which has previously been used to study gender bias in the delivery of
  job ads~\cite{Imana2021}.
However, researchers using BISG or other algorithms should re-evaluate
  error for the algorithm and location they study.

From the North Carolina voter dataset
  we take a sample of 100,000 individuals
  and apply BISG.
We then calculate the inference error rates
  over the sample population.
For real-world experiments, in general, these values will not be known,
  since any specific sample will vary randomly.
However, this simplified model can help demonstrate propagation of error.
Here we consider only expected error; we leave exploring variance of inference error as future work.

\subsection{Varying Audience Size}
   \label{sec:vary_sample_size}
  
 We first evaluate how the sample size available for auditing affects
  whether inference error leads to hiding skew that exists.
For this simulation,
  we vary the size of the targeted audience used for auditing ($|U|$)
  and compare the outcome of skew evaluation with and
  without correcting for inference error.
We fix the level of inference error to $\myV{Th}_{\Vs{BISG}} = 0.5$
  and $R=0.065$.
We fix the BISG threshold to $0.5$ to be consistent with
  VRS's implementation of BISG~\cite{metatech}.
$R=0.065$ also gives us the number of impressions per ad
  that is roughly consistent with real-world delivery rates from prior work~\cite{Imana2021}.
For example, using the definitions in \autoref{sec:setup_for_though_exp},
  at $R=0.065$, an audience of size of $10,000$ and $30,000$
  will result in approximately $650$ and $2,000$ impressions, respectively.

From the left column of \autoref{fig:vary_all_figures_combined},
  we find ignoring inference error leads to missing skew
   when the targeted audience is small.
 In each subfigure,
  the horizontal dotted line represents the threshold
  for statistical significance.
The red shaded region is the range of $S$ where
  we detect skew when targeting with true race
  (black dots are above the line) but we fail to detect
  the skew when targeting with inferred race but
  ignore inference error (red cross marks below the line).
If the skew we measure using our inference-aware method
  (green circles) is above the horizontal line in the shaded
  region, it indicates we successfully detect a skew
  that we would miss if we do not correct for inference error.

As shown in the red-shaded region in \autoref{fig:vary_all_figures_combined}(a),
  when $|U|=10,000$,
  ignoring inference error hides
  skew (red cross-marks below the horizontal line)
  when $S$ is in approximately in the range $\left[0.74, 0.92\right]$.
In the same range, we see \emph{our solution correctly
  detects skew} (green circles above the horizontal line).  
The width of the range decreases as we increase
  the audience size to $|U|=30,000$ and $|U|=60,000$,
  as shown in \autoref{fig:vary_all_figures_combined}(b) and
  \autoref{fig:vary_all_figures_combined}(c),
  respectively.
In \autoref{sec:vary_both_th_aud_size},
  we report the interval and width
  for larger audience sizes for which we see a similar trend.
 
 Therefore, \emph{using a larger audience can reduce error
   and increase confidence in skew detection}.
However,
  one must consider the size of the skew before reaching
  conclusions about harm.
 In addition, increasing the audience size can be costly
   in practical scenarios where access to data is limited,
  demonstrating our error correction is a valuable tool.
 
\subsection{Varying Level of Inference Error}
   \label{sec:vary_threshold}
   
We next evaluate how different levels of inference
  error rates lead to missing skew that exists and
  show how our proposed correction successfully detects the skew.
We vary inference error by setting the
  probability threshold
  we use for BISG
  ($\myV{Th}_{\Vs{BISG}}$),
  and compare the outcome of skew evaluation
  with and without correcting for inference error.
We fix $|U| = 30,000$ and $R=0.065$.

We vary inference error by setting
  $\myV{Th}_{\Vs{BISG}}$ to 0.5, 0.7, and 0.9.
We start at 0.5 because it is the threshold used
  by Meta's VRS~\cite{metatech}, and evaluate how
  skew detection improves as we increase the threshold.
Each part of the figure uses a specific
  BISG threshold.
We fix the targeted audience size to $|U|=30,000$ and
  $R=0.065$.
As the threshold increases, the corresponding error rates
  either decrease or stay the same.
For example,
  $\myV{FDR}_{b,a}$ decreases from 0.47 to 0.38 when
  the threshold changes from 0.5 to 0.7,
  respectively.

We find inference error is more likely to hide skew that exits
  when we use a lower threshold
  (inference error is higher),
  as shown in the right column of \autoref{fig:vary_all_figures_combined}.  
As shown in \autoref{fig:vary_all_figures_combined}(d),
 the width of the shaded region where we miss skew
  is the largest when
  $\myV{Th}_{\Vs{BISG}} = 0.5$.
\autoref{fig:vary_all_figures_combined}(e) and \autoref{fig:vary_all_figures_combined}(f)
  show that the width decreases as we increase
  the threshold to 0.7 and 0.9.
In the shaded regions,
  we see our inference-aware evaluation (green circles)
  are above the horizontal line,
  showing \emph{we successfully detect skew}.

Our results suggest we should also \emph{use a higher inference threshold
  when possible}.
A higher threshold will increases confidence in the results,
  but it does so by excluding individuals with names that are not strongly 
  correlated with race,
  so it also increases costs in audience construction.
Therefore, we must ensure the delivery audience
  is large enough for statistical analysis after excluding
  individuals that receive a low score.

\section{Future Work}
\reviewfix{}
Our methodology addresses a key challenge in applying
  the paired-ads approach to audit ad delivery algorithms
  when demographic attributes must be inferred.
However, several limitations remain that point to directions
  for future work.
First, our correction method models only the expected inference error
  and does not capture variance.
In particular, we do not account for uncertainty in the estimation
  of error rates themselves, or how such uncertainty propagates
  to the final statistical test of delivery skew. Incorporating variance in
  error rates in an area future work.

Second, our model
  focuses on 
  relationships between two demographic
  groups (designated A and B),
  and places other groups in the ``Other'' category.
Generalizing the method to support relationships between more than two gropus
  would expand the applicability of our approach.

Third, we assume that skew in delivery operates on
  true demographic attributes and is independent of inference error.
This assumption simplifies the analysis and is partially supported by existing implementations, such as Meta's reliance on user embeddings that
  correlate with true race.
Exploring dependence between algorithmic skew and demographic inference uncertainty is left to future work.

Finally, while we validate our inference-aware method
  using simulations, future work could explore empirical
  validations in real-world settings.

\section{Conclusion}
We have shown the importance of considering 
  error in inferred demographics during black-box audits of ad delivery algorithms
  for bias.
Our proposal applies attribute inference to the unique setting of the paired-ads method
  that remains a key black-box auditing tool for keeping platforms accountable.
This application differs from prior studies that consider
  inference error when training classifiers.
Our analysis relies on only aggregate statistics,
  making it applicable to black-box evaluation of ad delivery bias.
This approach differs from previous work
  that required
  individual-level data and, therefore, was not applicable to black-box evaluation.
By showing how to \emph{account} for inference error,
  we expand the auditing toolkit,
  allowing evaluation of bias in
  ad delivery algorithms 
  for domains
  where demographic attributes are inaccessible. 
Our approach can be extended to audit disparities
  across protected attributes beyond race, such as gender or age,
  for which the inference methods can be developed and error rates can be
  estimated using datasets with ground-truth demographics.


\begin{acks}
This work was funded in part by 
  National Science Foundation grants
  CNS-1956435, %
  CNS-2344925, %
and
  CNS-2319409. %
We thank our reviewers for helpful feedback.
\end{acks}

\section*{Ethics}

Our analysis is ethically sound because it 
  provides a better understanding of how to use inferred
  demographic attributes for auditing ad delivery algorithms for bias,
  with consideration of inference error and without creating new privacy
  risks to individuals.
Overall, our work has a positive outcome by improving our understanding
  of how to audit social media algorithms, and important part
  of today's Internet.
It poses minimal risks for several reasons:

Our work poses no new privacy risks to individuals because
  our input data is currently publicly available in existing voter datasets.
Additionally, our approach does not involve collecting
  individual-level identifiers as platforms report
  only aggregate information about ad recipients.
In cases where the GDPR applies, inferred demographic
  attributes may be considered
  personal data for which consent is required~\cite{Bogen2024}.
We do not solicit informed consent
  because we have no means to directly interact with the individuals 
  in our data.

The methodology we adopt from prior work
  (using paired ads~\cite{Ali2019a, Imana2021, Imana2024}) poses minimal cost on individuals.
The ad budget is small (\$50 per ad campaign) and
  has minimal influence on the overall on-line ad market
  or on the full mix of ads any individual sees.
In addition, we consider ads that link to real-world
  economic opportunities;
  which are of potential benefit to the recipients.

\bibliographystyle{ACM-Reference-Format}
%
%
%

%
%
%
%
%
%
%
%
%
%

%


\appendix


\section{Proofs: Theorem \ref{thm:error_no_effect} and Theorem \ref{thm:error_underestimate_skew}}
  \label{sec:appendix_proof_inf_error_effect}
%
%
%
%

\textbf{Proof of \autoref{thm:error_no_effect}}: In this proof,
  we show that inference error has no effect on auditing if there is no true underlying skew
  in the ad delivery algorithm.

\begin{proof}
We derive and compare the skew that an auditor measures
  when the targets audience $U$ is constructed using
  true and inferred demographic attributes.
  
First, we first consider a case where $U$ is constructed using true race.
Based on our setup in~\autoref{sec:setup_notations}, we know $D_t = s_{2,a,t} -  s_{1,a,t}$, where
$s_{1,a,t} = \frac{{n_{1,a,t}}}{ n_{1,a,t} + n_{1,b,t}}  \hspace{0.5em} s_{2,a,t} = \frac{{n_{2,a,t}}}{n_{2,a,t} + n_{2,b,t}}$.

The platform's report
  the number of individuals in group A and group B
  that see each of the ads
  ($n_{1,a,t}$, $n_{1,b,t}$, $n_{2,a,t}$, $n_{2,b,t}$)
  individuals that see the ad.
We can write these quantities
 using the demographic composition of the targeted audience ($u_{a,t}$ and $u_{b,t}$),
  the delivery rate ($R$) and the skew in the ad delivery algorithm ($S$):

\begin{equation}\label{eq:del_aud_true_race}  
\begin{aligned}
n_{1,a,t} & = u_{a,t} \times R \times S \\
n_{2,a,t} & = u_{a,t} \times R \\
n_{1,b,t} & = u_{b,t} \times R \times (2-S)\\
n_{2,b,t} & = u_{b,t} \times R \\
\end{aligned}
\end{equation}

Using the fact that our audience $U$ contains equal number individuals from both groups A and B: $u_{a,t} = u_{b,t} = \frac{|U|}{2}$, we can now derive the fraction of individuals whose true membership is in group A that see each ad as follows:

\begin{equation}\label{eq:del_aud_size}  
\begin{aligned}
 s_{1,a,t} = \frac{{n_{1,a,t}}}{ n_{1,a,t} + n_{1,b,t}} &= \frac{u_{a,t} \times R \times S}{(u_{a,t} \times R \times S)+(u_{b,t} \times R \times (2-S))} \\
 &=  \frac{u_{a,t} \times S}{(u_{a,t} \times S)+(u_{b,t} \times (2-S))} \\
 &=  \frac{\frac{|U|}{2} \times S}{(\frac{|U|}{2} \times S)+(\frac{|U|}{2} \times (2-S))} \\
  &=  \frac{S}{S+(2-S)} = \frac{S}{2} \\
 s_{2,a,t} = \frac{{n_{2,a,t}}}{ n_{2,a,t} + n_{2,b,t}} &= \frac{u_{a,t} \times R}{(u_{a,t} \times R)+(u_{b,t} \times R)} \\
 &= \frac{u_{a,t}}{(u_{a,t})+(u_{b,t} )} \\
 &= \frac{\frac{|U|}{2}}{\frac{|U|}{2}+\frac{|U|}{2}} = \frac{1}{2}
\end{aligned}
\end{equation}

Plugging in $S=1$, $D_t = s_{2,a,t} - s_{1,a,t} =  \frac{1}{2} - \frac{S}{2} =  \frac{1}{2} - \frac{1}{2} = 0$.

We next consider the case where an auditor targets an audience $U$ constructed using inferred race and show $D_i= 0$.

In this case, the report from the platform gives us
  the number of individuals that see each of the ads
  whose are \emph{inferred} to be in group A or B
  ($n_{1,a,i}$, $n_{1,b,i}$, $n_{2,a,i}$, $n_{2,b,i}$).
We can write these quantities
  using the racial composition of the targeted audience ($u_{a,t}$ and $u_{b,t}$),
  the error rates of the attribute inference method,
  the delivery rate ($R$) and the skew in the ad delivery algorithm $S$:
  
\begin{equation}\label{eq:del_aud_inf_race}  
\begin{aligned}
  n_{1,a,i} & = u_{a,i} \times (1-\myV{FDR}_{*,a}) \times R \times S \\
  & + u_{a,i} \times \myV{FDR}_{b,a} \times R \times (2-S) \\
  & + u_{a,i} \times \myV{FDR}_{o,a} \times R \times (2-S) \\ 
  & = R \times u_{a,i} \left[S \times  (1-\myV{FDR}_{*,a}) + (2-S) \times (\myV{FDR}_{b,a} + \myV{FDR}_{o,a}) \right]  \\
  & = R \times u_{a,i} \left[S \times (1-\myV{FDR}_{*,a}) +  (2-S) \times (\myV{FDR}_{*,a}) \right] \\
\end{aligned}
\end{equation}

\begin{equation}\label{eq:del_aud_inf_race2}
\begin{aligned}
n_{1,b,i} &= u_{b,i} \times \myV{FDR}_{a,b} \times R \times S  \\
&+ u_{b,i} \times (1-\myV{FDR}_{*,b}) \times R \times (2-S)  \\
&+ u_{b,i} \times \myV{FDR}_{o,b} \times R \times (2-S) \\
& = R \times u_{b,i} \left[S \times  (\myV{FDR}_{a,b}) + (2-S) \times (1 - \myV{FDR}_{*,b} + \myV{FDR}_{o,b}) \right] \\
& = R \times u_{b,i} \left[S \times  (\myV{FDR}_{a,b}) + (2-S) \times (1 - \myV{FDR}_{a,b} \right]
\end{aligned}
\end{equation}

We can now derive the fraction of individuals inferred to be in group A that see each ad as follows.
From our setup, we know that the audience $U$ contains equal number of individuals inferred as group A and B: $u_{a,i} = u_{b,i} = \frac{|U|}{2}$.

\begin{equation}\label{eq:del_aud_size_inf_race}
\begin{aligned}
 s_{1,a,i} & = \frac{{n_{1,a,i}}}{ n_{1,a,i} + n_{1,b,i}} \\
 &= \frac{\left[S \times (1-\myV{FDR}_{*,a}) +  (2-S) \times (\myV{FDR}_{*,a}) \right]}{ \parbox{2.3in}{$\left[S \times (1-\myV{FDR}_{*,a}) +  (2-S) \times (\myV{FDR}_{*,a}) \right]$ \\
 $+ \left[S \times (\myV{FDR}_{a,b}) + (2-S) \times (1 - \myV{FDR}_{a,b} \right]$}} \\
  s_{2,a,i} & = \frac{{n_{2,a,i}}}{ n_{2,a,i} + n_{2,b,i}} \\
 &= \frac{(1-\myV{FDR}_{*,a}) + (\myV{FDR}_{*,a}) }{(1-\myV{FDR}_{*,a}) +  (\myV{FDR}_{*,a}) + (\myV{FDR}_{a,b}) + (1 - \myV{FDR}_{a,b}) } \\
&= \frac{1}{1 + 1} = \frac{1}{2}
\end{aligned}
\end{equation}

 Plugging in $S=1$:
\begin{equation*}  
\begin{aligned}
s_{1,a,i} &= \frac{(1-\myV{FDR}_{*,a}) +  \myV{FDR}_{*,a} }{(1-\myV{FDR}_{*,a}) +  \myV{FDR}_{*,a}  + (\myV{FDR}_{a,b}) + (1 - \myV{FDR}_{a,b}) } \\
&= \frac{1}{1 + 1} = \frac{1}{2}
\end{aligned}
\end{equation*}

Therefore, $D_i = s_{2,a,i} - s_{1,a,i} =  \frac{1}{2} - \frac{1}{2} = 0 = D_t$.

\end{proof}

\textbf{Proof of \autoref{thm:error_underestimate_skew}:} We next present our proof for \autoref{thm:error_underestimate_skew}
  that shows ignoring inference error leads to underestimating skew.
  
\begin{proof}
To show $|D_i| < |D_t|$, we consider two cases:
  when $D_t > 0$, we show $D_i < D_t $;
  when $D_t < 0$, we show $D_i > D_t$.

\textbf{Case 1 ($D_t > 0$)}:
  we would like to show $D_i < D_t$.
Plugging the definition of $D$ in \autoref{sec:setup_notations}, we want to show: 
$s_{2,a,t} -  s_{1,a,t} > s_{2,a,i} -  s_{1,a,i}$.
From \autoref{eq:del_aud_size} and \autoref{eq:del_aud_size_inf_race}, we know $s_{2,a,t} = s_{2,a,i} = \frac{1}{2}$, so they cancel out.
We are then left with showing $s_{1,a,i} - s_{1,a,t} > 0$.
Plugging in the expressions we derived in \autoref{eq:del_aud_size} and \autoref{eq:del_aud_size_inf_race}:

\begin{equation*}  
\begin{aligned}
s_{1,a,i} - s_{1,a,t} = \left( \frac{\left[S \times (1-\myV{FDR}_{*,a}) +  (2-S) \times (\myV{FDR}_{*,a}) \right]}{ \parbox{2.1in}{$\left[S \times (1-\myV{FDR}_{*,a}) +  (2-S) \times (\myV{FDR}_{*,a}) \right]$ \\
 $+ \left[S \times (\myV{FDR}_{a,b}) + (2-S) \times (1 - \myV{FDR}_{a,b}) \right]$}} \right) - \frac{S}{2} > 0
\end{aligned}
\end{equation*}

For brevity, we define the following symbols: $p = \myV{FDR}_{*,a}$ and $q= \myV{FDR}_{a,b}$. After rearranging the terms, the inequality is then given by:
\begin{equation}\label{eq:simplified_ineq_case1}
\begin{aligned}
\frac{(S-1) \times ((p \times (S-2)) - q \times S)}{(p \times (S-1)) + (q \times (-S)) + (q-1)} < 0
\end{aligned}
\end{equation}

To show the above inequality holds, we use the following:
\begin{equation}\label{eq:proof_param_ranges}
\begin{aligned}
 &\text{$0 \le S < 1$ because $D_t > 0$} \\
 &\text{$0 < p \le 1$ by definition of $\myV{FDR}_{*,a}$} \\
 &\text{$0 < q \le 1$ by definition of $\myV{FDR}_{a,b}$} \\
\end{aligned}
\end{equation}

We prove \autoref{eq:simplified_ineq_case1},
  by show the numerator and denominator have opposite signs,
  because the numerator is positive 
  and the denominator is negative for all possible values of $S$, $p$ and $q$.

\textbf{Case 1a}:  
We show $(S-1) \times ((p \times (S-2)) - q \times S)$ is positive
  because both terms $(S-1)$ and $((p \times (S-2)) - q \times S$ are negative.
For the first term, 
  $S -1 < 0$ because, by \autoref{eq:proof_param_ranges}, we know $0 \le S < 1$.
For the second term, $p \times (S-2) < 0$ because $p > 0$ and $S-2 < 0$.
Finally, $q \times S \ge 0$ by \autoref{eq:proof_param_ranges}.
Since we subtract a non-negative term from a negative term, 
  the result is always negative.

\textbf{Case 1b}: we show the denominator
  $(p \times (S-1)) + (q \times (-S)) + (q-1)$ is negative.
We can rearrange the expression and show
  $(p-q) (S-1) < 1$. From \autoref{eq:proof_param_ranges},
  we know, for the first term,
  $-1 < p - q < 1$,
  and for the second term,
  $-1 \le S - 1 < 0$.
Therefore, the product of the two terms 
  $(p-q) (S-1) < 1$ holds.

Therefore, we have shown \autoref{eq:simplified_ineq_case1} holds because both the numerator and denominator have opposite signs.

\textbf{Case 2 ($D_t < 0$)}: we would like to show $D_i > D_t$. Following the same steps we took for Case 1,
  we can derive the following inequality where the only difference from \autoref{eq:simplified_ineq_case1} is the direction of the inequality:
\begin{equation}\label{eq:simplified_ineq_case2}
\begin{aligned}
\frac{(S-1) \times ((p \times (S-2)) - q \times S)}{(p \times (S-1)) + (q \times (-S)) + (q-1)} > 0
\end{aligned}
\end{equation}

For this case, we know the following:
\begin{equation}\label{eq:proof_param_ranges_case2}
\begin{aligned}
 &\text{$1 < S \le 2$ because $D_t < 0$} \\
 &\text{$0 < p \le 1$ by definition of $\myV{FDR}_{*,a}$} \\
 &\text{$0 < q \le 1$ by definition of $\myV{FDR}_{a,b}$} \\
\end{aligned}
\end{equation}

To prove \autoref{eq:simplified_ineq_case2}, it suffices to show both the denominator and numerator are negative for all possible values of $S$, $p$ and $q$. From Case 1,
  we already know the denominator is negative, so we just show the numerator is negative.

To show the numerator
  $(S-1) \times ((p \times (S-2)) - q \times S)$ is negative,
  we need to shown the terms
    $(S-1)$ and
    $((p \times (S-2)) - q \times S)$
    have opposite signs.
By \autoref{eq:proof_param_ranges_case2},
  we know the first term $S - 1 > 0$,
  so we need to show the other term is negative.
For the other term,
  we know $(p \times (S-2)) \le 0$ because $p > 0$ and $S-2 \le 0$.
We also know $q \times S > 0$ by \autoref{eq:proof_param_ranges_case2}.
Therefore, we are subtracting a positive number ($q \times S$) from a number that is either negative or 0 ($p \times (S-2))$, resulting a negative term.

Therefore, $|D_i| < |D_t|$ holds for both cases.
\end{proof}   

\section{Thought Experiments on the Effects of Inferred Attributes}
  \label{sec:infer_race}

In \autoref{sec:infer_race_proofs}, we gave theoretical results
  that show the effect of attribute estimation on auditing ad delivery.
In this section,
  we explore thought experiments that provide concrete
  examples of how inference error affects the conclusions of
  an audit.
For concreteness,
  we use race as an example
  where group A and B defined in our notations (\autoref{sec:setup_notations})
  represent ``Black'' and ``White'' racial groups, respectively.

In our first thought-experiment,  there is no algorithmic skew (\autoref{thm:error_no_effect}).
We use this case to explore perspectives
  on what is true, what can be observed,
  and where they differ,
  to show how inference can potentially affect the conclusion.
In this baseline example, we find race inference error
  does not affect our evaluation of skew because
  there is no skew in the ad delivery algorithm.

In our second thought-experiment
  we add known algorithmic skew (\autoref{thm:error_underestimate_skew}).
We show we can detect this skew with statistical
  rigor given construct with true race.
We then show that when we infer race,
  inference error can hide the skew.
We lose significance because,
  although we intended for audiences with the same ratio of races
  ($u_{a,i}$ : $u_{b,i}$ is the same as $u_{a,t}$ : $u_{b,t}$),
  inference error means the \emph{actual} audience is different than expected.
We find this difference occurs when the inference method performs
  more poorly for Black individuals than White individuals
  ($\myV{FDR}_{*,a} > \myV{FDR}_{*,b}$),
  resulting in underrepresentation of Black individuals.
  
\subsection{Setup and Assumptions}

For our thought experiments,
  we assume we have omniscient knowledge both the true and inferred race of ad recipients;
  a luxury we lack in real applications.
We use omniscient information to compute a \emph{corrected} version
  of the outcome using an inferred population.
This corrected version lets us separate
  inference error from potential platform-induced skew.  
Because the skew in the ad delivery algorithm is fixed in each thought experiment,
  we expect the skew we measure when targeting with true race
  to be roughly the same as the skew we measure when
  targeting with inferred race but compute skew with our
  omniscient knowledge of true race.  
We use different notation to distinguish when the attribute
  is known, inferred or an omniscient as we show in \autoref{tab:delivery_notation_with_correction}.

We introduce notation for the corrected statistics we compute using our
  omniscient knowledge of true race when targeting with inferred race.
Let $n_{1,a,c}$ and $n_{1,b,c}$ represent the number of true Blacks and true Whites that saw the first ad.
Let  $n_{1,c}$ = $n_{1,a,c}$ + $n_{1,b,c}$.
We define $n_{2,c}$, $n_{2,a,c}$ and $n_{2,b,c}$ similarly for the second ad.
We do not include in $n_{1,c}$ and $n_{2,c}$ people of ``Other''
   race that may have seen an ad because our goal is to compute
   an estimate of the true statistics we would have computed
   if we targeted with true race.
By applying \autoref{eq:generic_Z_formula},
 the test statistic is given by:
\begin{equation}\label{eq:error_Z_formula_corrected}
Z_c = \myV{ZTEST}(n_{1,a,c}, n_{1,b,c}, n_{2,a,c}, n_{2,b,c})
\end{equation}

For both thought experiments,
  we use the inference error values we observe by applying BISG to North Carolina
  voter data
  using $\myV{Th}_{\Vs{BISG}} = 0.5$:
  $\myV{FDR}_{a,b}=0.14$, $\myV{FDR}_{o,b}=0.03$, $\myV{FDR}_{b,a}=0.47$, $\myV{FDR}_{o,a}=0.03$ (see \autoref{sec:validation_setup} for details).
 
When operating with true race,
  we calculate the test statistic for statistical significance using $Z_{\myV{t}}$.
When operating with inferred race
  we calculate the test statistic for statistical significance using $Z_{\myV{i}}$.
All statistical tests are conducted at a significance level of $\alpha = 0.05$.
We conclude statistical significance when a statistic is above the critical value $Z_{\alpha} = 1.64$.

\subsection{Example Without Algorithmic Skew}
   \label{sec:exp_1_sec}

    \begin{table*}
    \centering
    \begin{tabular}{c|c|c|cc}
    \hline
    \makecell{Parameters \\ for this \\ example} & \multicolumn{3}{c}{\makecell[l]{
    |U|=30,000, R=0.065, \textcolor{red}{S=1}
     \\ 
    Inference error rates: \\
    $FDR_{a,b}$=0.14,
    $FDR_{o,b}$=0.03, \\
    $FDR_{b,a}$=0.47, 
    $FDR_{o,a}$=0.03 \\
    Group A: Blacks; Group B: White
    }}\hspace*{1in} \\
    \hline
    \hline
    \makecell{Targeted \\ using} & \makecell{True attribute} & \makecell{Inferred attribute} & \makecell{Inferred attribute  \\ (omniscient  \\ correction)}\\
    \hline
    \makecell{Targeted \\ audience} & \makecell[l]{
    |U|={30,000} \\
    \hspace{3.2em}\textcolor{black}{[100\%]} \\
    \hspace{0.2em}\textcolor{black}{$u_{a,t}$=15,000} \\
    \hspace{3.2em}\textcolor{black}{[50\%]} \\
    \hspace{0.2em}\textcolor{black}{$u_{b,t}$=15,000} \\
    \hspace{3.2em}\textcolor{black}{[50\%]} \\
    } & \makecell[l]{
    |U|={30,000} \\
    \hspace{0.2em}$u_{a,i}$=15,000 [100\%] \\
    \hspace{0.4em}\textcolor{black}{7,466} A [49.8\%]\\
    \hspace{0.4em}\textcolor{black}{7,090} B [47.3\%] \\
    \hspace{0.4em}\textcolor{black}{444} O [3.0\%] \\
    \hspace{0.2em}$u_{b,i}$=15,000 [100\%] \\
    \hspace{0.4em}\textcolor{black}{2,156} A [14.4\%]\\
    \hspace{0.4em}\textcolor{black}{12,369} B [82.5\%]\\
    \hspace{0.4em}\textcolor{black}{476} O [3.2\%]\\
    } & \makecell[l]{
    {29,080} (A+B) \\
    \hspace{0.4em}$u_{a,c}$ \\ \hspace{0.4em}=7,466+2,156 \\ \hspace{0.4em}=9,621 A [33.1\%] \\ \\
    \hspace{0.4em}$u_{b,c}$ \\ \hspace{0.4em}=7,090+12,369 \\ \hspace{0.4em}=19,460 B [66.9\%]
    }  \\
    \hline
    \hline
    \makecell{Delivery \\ audience \\ for ad 1:  \\ rate of $R \times S$ \\ for group A; \\ $R \times (2-S)$ \\ for group B and O}
     & \makecell[l]{
    $n_{1,t}$=1,950 \\
    \hspace{3.2em}\textcolor{black}{[100\%]} \\
    \hspace{0.2em}\textcolor{black}{$n_{1,a,t}$=975} \\
    \hspace{3.2em}\textcolor{black}{[50\%]} \\
    \hspace{0.2em}\textcolor{black}{$n_{1,b,t}$=975} \\
    \hspace{3.2em}\textcolor{black}{[50\%]} \\
    } & \makecell[l]{
    $n_{1,i}$=1,950 \\
    \hspace{0.2em}$n_{1,a,i}$=975 [100\%]  \\
    \hspace{0.4em}\textcolor{black}{485} A [49.8\%]\\
    \hspace{0.4em}\textcolor{black}{461} B [47.3\%]\\
    \hspace{0.4em}\textcolor{black}{29} O [3.0\%]\\
    \hspace{0.2em}$n_{1,b,i}$=975 [100\%] \\
    \hspace{0.4em}\textcolor{black}{140} A [14.4\%]\\
    \hspace{0.4em}\textcolor{black}{804} B [82.5\%]\\
    \hspace{0.4em}\textcolor{black}{31} O [3.2\%]\\
    } & \makecell[l]{
    $n_{1,c}$=1,890 (A+B) \\
    \hspace{0.4em}$n_{1,a,c}$ \\ \hspace{0.4em}=485+140 \\ \hspace{0.4em}=625 A [33.1\%] \\ \\
    \hspace{0.4em}$n_{1,b,c}$ \\ \hspace{0.4em}=461+804 \\ \hspace{0.4em}=1,265 B [66.9\%] 
    } \\
    \hline
    \makecell{Delivery \\ audience \\ for ad 2: \\ rate of $R$ \\ for all} & \makecell[l]{
    $n_{2,t}$=1,950 \\
    \hspace{3.2em}\textcolor{black}{[100\%]} \\
    \hspace{0.2em}\textcolor{black}{$n_{2,a,t}$=975} \\
    \hspace{3.2em}\textcolor{black}{[50\%]} \\
    \hspace{0.2em}\textcolor{black}{$n_{2,b,t}$=975} \\
    \hspace{3.2em}\textcolor{black}{[50\%]} \\
    } & \makecell[l]{
    $n_{2,i}$=1,950 \\
    \hspace{0.2em}$n_{2,a,i}$=975 [100\%] \\
    \hspace{0.4em}\textcolor{black}{485} A [49.8\%]\\
    \hspace{0.4em}\textcolor{black}{461} B [47.3\%]\\
    \hspace{0.4em}\textcolor{black}{29} O [3.0\%]\\
    \hspace{0.2em}$n_{2,b,i}$=975 [100\%] \\
    \hspace{0.4em}\textcolor{black}{140} A [14.4\%]\\
    \hspace{0.4em}\textcolor{black}{804} B [82.5\%]\\
    \hspace{0.4em}\textcolor{black}{31} O [3.2\%]\\
    } & \makecell[l]{
    $n_{2,c}$=1,890 (A+B) \\
    \hspace{0.4em}$n_{2,a,c}$ \\ \hspace{0.4em}=485+140 \\ \hspace{0.4em}=625 A [33.1\%] \\ \\
    \hspace{0.4em}$n_{2,b,c}$ \\ \hspace{0.4em}=461+804 \\ \hspace{0.4em}=1,265 B [66.9\%]
    } \\
    \hline
    \hline
    \makecell{Skew \\ evaluation} & \makecell{
    $s_{1,a,t} = 0.50$ \\
    $s_{2,a,t} = 0.50$ \\
    $D_t = 0.00$
    \\
    $Z_{\myV{t}} = 0.00$ \\
    ($\leq$ 1.64) \\ (Not signif.)
    }  & \makecell{
    $s_{1,a,i} = 0.50$ \\
    $s_{2,a,i} = 0.50$ \\
    $D_i = 0.00$
    \\
    $Z_i = 0.00$ \\
    ($\leq$ 1.64) \\ (Not signif.)
    } & \makecell{
    $s_{1,a,c} = 0.33$ \\
    $s_{2,a,c} = 0.33$ \\
    $D_c = 0.00$
    \\
    $Z_{\myV{i,c}} = 0.00$ \\
    ($\leq$ 1.64) \\ (Not signif.)
    } \\
    \hline
    \end{tabular}
    \caption{A baseline example where race inference error does not affect evaluation of skew in ad delivery because there is no skew in the platform's ad delivery algorithm ($S = 1$).}
    \label{tab:ex_baseline_corrected}
    \end{table*}
    
We first consider the case with no skew for either ad ($S=1$).
Here we evaluate how inference error changes the target audience,
  slightly distorting the results, although not the conclusion.

We work through the example in 
  \autoref{tab:ex_baseline_corrected}.
The left column shows a case where we construct our target audience
  using true race.
We begin with an audience of 30,000 people (top row, $u_{a,t}=u_{b,t}=15,000$).
Each ad is delivered to $R=0.065$ of them ($n_{1,a,t}=n_{1,b,t}=n_{2,a,t}=n_{2,b,t}=975$).
Since there is no skew in the ad delivery algorithm,
  both ads are delivered to the same
  fraction of true Blacks ($s_{1,a,t} = s_{2,a,t} = 0.5$),
  the test statistics are zero  ($D_t = 0$; $Z_t = 0$),
  and we conclude there is no statistically significant skew.
Using the method established in \autoref{sec:setup_notations},
  and without inference,
  we correctly confirm there is no skew.

Now we analyze what delivery statistics we observe
  when we target using \emph{inferred} race,
  in the middle column.
Again, we choose 30,000 targets,
  and we \emph{think} they are 15,000/15,000 Black/White from inference.
However, this our equal mix is not realized---because of inference error,
  more than half of our Black targets (50.3\%= 47.3\%+3.0\%) are not actually Black,
  while only 17.6\% of our White targets are mis-identified.
In this secnario with no skew,
  delivery results for both ads are identical to the mix of the target audience
  (both audiences are multiplied by delivery rate $R$, and $S=1$).

We evaluate the results when the audiences are targeted using inferred race
  in two ways:
  in the middle column, we consider what we can tell based on inferred race only.
In the third column, we examine who \emph{really} saw the ad,
    computing statistics with our omniscient knowledge of true race.
To compute the third column, 
  we take the audience we use, 
  which we believe is an equal 50/50\% split by race.
We compute how much of this inference was incorrect,
  to find the actual audience we selected.
We see it has only 9,621 Blacks (33\%)
  and 19,460 Whites (67\%).
We conclude that,
  \emph{while we thought we had a 50/50\% audience,
  we actually have a 33\%/67\% audience,
  where Blacks are represented much less than we expect.}
With omniscient true race,
  the fraction of Blacks that see the first ad is $s_{1,a,c}$=0.33,
  showing the under-presentation of Blacks in the 33/67\% audience
   propagates to ad delivery.
 However, using only inferred information (middle column)
   hides this fact: $s_{1,a,i} = 0.5$.
Fortunately, the same case is true for both ads,  as we assume that there is no skew, $S=1$,
  resulting in no net effect that alters our conclusion,
 so our evaluation of skew is correct
  regardless of inference error.

\subsection{Example Where Algorithmic Skew is Hidden}
  \label{sec:exp_2_sec}

We now inject a known amount of skew ($S$)
  into ad delivery
  for our second example in \autoref{tab:ex_miss_skew_corrected}.
We use a specific value of $S=0.87$ to provide a concrete example;
  we study a range of values in~\autoref{sec:validation_summary}.
In this scenario, we will show inference error during target audience creation
  hides algorithm skew.
It gives us an incorrect conclusion,
  and a different outcome than had we tested with correct information about race.

Left column labelled ``true race'' in \autoref{tab:ex_miss_skew_corrected}
  shows auditing using perfect information about race.
We again target an audience of 30,000
  (omitted in \autoref{tab:ex_miss_skew_corrected}, but the
  same as the top row of \autoref{tab:ex_baseline_corrected}).
While ad 2 is delivered equally by race (975 each for Blacks and Whites,
  the same as the third row of \autoref{tab:ex_baseline_corrected}),
  delivery of ad 1 is skewed by the platform ($S=0.87$),
  going to more Whites than Blacks
    (1,102 vs. 848, as shown in \autoref{tab:ex_miss_skew_corrected}).
This difference appears in the fraction of
  impressions of the first and second ad seen by Blacks,
  with the delivery audience of the first ad ($s_{1,a,t} = 0.43$) having a larger fraction of Blacks than the delivery audience of the second ad ($s_{2,a,t} = 0.5$).
The relative difference in delivery for the two ads for Blacks ($D_t=0.07$) 
  and the test statistic of $Z_t = 4.07$
  show a statistically significant skew in the ad delivery algorithm.  
This thought experiment with known race data correctly demonstrates 
  one can prove skew exists,
  as we expect from \autoref{sec:setup_notations}.

    \begin{table*}
    \centering
    \begin{tabular}{c|c|c|c}
    \hline
    \makecell{Parameters \\ for this \\ example} & \multicolumn{3}{c}{\makecell[l]{
    |U|=30,000, R=0.065, \textcolor{red}{S=0.87}
     \\ 
    Inference error rates: \\
    $FDR_{a,b}$=0.14,
    $FDR_{o,b}$=0.03, \\
    $FDR_{b,a}$=0.47, 
    $FDR_{o,a}$=0.03 \\
    Group A: Blacks; Group B: White
    }}\hspace*{1in} \\
    \hline
    \hline
    \makecell{Targeted \\ using} & \makecell{True attribute} & \makecell{Inferred attribute} & \makecell{Inferred attribute  \\ (omniscient  \\ correction)}\\
    \hline
\multicolumn{4}{c}{Targeted audience: same as the top row of \autoref{tab:ex_baseline_corrected}} \\
    \hline
    \hline
    \makecell{Delivery \\ audience \\ for ad 1:  \\ rate of $R \times S$ \\ for group A; \\ $R \times (2-S)$ \\ for group B and O}
     & \makecell[l]{
    $n_{1,t}$=1,950 \\
    \hspace{3.2em}\textcolor{black}{[100\%]} \\
    \hspace{0.2em}\textcolor{black}{$n_{1,a,t}$=848} \\
    \hspace{3.2em}\textcolor{black}{[44\%]} \\
    \hspace{0.2em}\textcolor{black}{$n_{1,b,t}$=1,102} \\
    \hspace{3.2em}\textcolor{black}{[56\%]} \\
    } & \makecell[l]{
    $n_{1,i}$=2,041 \\
    \hspace{0.2em}$n_{1,a,i}$=976 [100\%]  \\
    \hspace{0.4em}\textcolor{black}{422} A [43.3\%]\\
    \hspace{0.4em}\textcolor{black}{521} B [53.4\%]\\
    \hspace{0.4em}\textcolor{black}{33} O [3.3\%]\\
    \hspace{0.2em}$n_{1,b,i}$=1,065 [100\%] \\
    \hspace{0.4em}\textcolor{black}{122} A [11.4\%]\\
    \hspace{0.4em}\textcolor{black}{909} B [85.3\%]\\
    \hspace{0.4em}\textcolor{black}{35} O [3.3\%]\\
    } & \makecell[l]{
    $n_{1,c}$=1,973 (A+B) \\
    \hspace{0.4em}$n_{1,a,c}$ \\ \hspace{0.4em}=422+122 \\ \hspace{0.4em}=544 A [27.6\%] \\ \\
    \hspace{0.4em}$n_{1,b,c}$ \\ \hspace{0.4em}=521+909 \\ \hspace{0.4em}=1,429 B [72.4\%] 
    } \\
    \hline
\multicolumn{4}{c}{Delivery audience for ad 2: same as the penultimate row of \autoref{tab:ex_baseline_corrected}} \\
    \hline 
    \hline
    \makecell{Skew \\ evaluation} & \makecell{
    $s_{1,a,t} = 0.43$ \\
    $s_{2,a,t} = 0.50$ \\
    $D_t = 0.07$
    \\
    $Z_{\myV{t}} = 4.07$ \\
    (> 1.64) \\ (Stat. sign.)
    }  & \makecell{
    $s_{1,a,i} = 0.48$ \\
    $s_{2,a,i} = 0.50$ \\
    $D_i = 0.02$
    \\
    $Z_i = 1.39$ \\
    (<= 1.64) \\ (Not signif.)
    } & \makecell{
    $s_{1,a,c} = 0.28$ \\
    $s_{2,a,c} = 0.33$ \\
    $D_c = 0.06$
    \\
    $Z_{\myV{i,c}} = 3.73$ \\
    (> 1.64) \\ (Stat. sign.)
    } \\
    \hline
    \end{tabular}
    \caption{An example with known inference error demonstrating that inference of the target audience can result in missing detection of actual skew. Compared to the baseline example (\autoref{tab:ex_baseline_corrected}), the only change is to inject known skew ($S = 0.87$), which affects the delivery of ad 1.}
    \label{tab:ex_miss_skew_corrected}
    \end{table*}

 In the middle column, we analyze delivery statistics observed
  when targeting using \emph{inferred} race.
We know algorithmic skew exists,
  so we examine how inference changes our evaluation,
  because the impressions ad 1 receives
  are influenced by two factors:
  algorithmic skew 
  \emph{and} the unexpected racial mix in the audience.
Recall that the platform's algorithms propagate bias according to
   true race,
  since the platform's algorithms use rich data
  and it does not use our inferred race, 
  thus a different audience mix could confuse 
  our evaluation for algorithm skew.
These factors have several results on our analysis:
First, %
  the skew we measure using inferred race ($D_i=0.02$)
  \emph{underestimates} the
  true skew we measured when targeting using true race ($D_t=0.07$).
This underestimation
  prevents statistically significant detection of skew according to
  inferred race ($Z_{\myV{i}} = 1.39$).
In this case, \emph{inference error in audience construction
  results in being unable to prove skew exists (with statistical significance),
  even though we know it exists.}
%
%

%
%
%
%
%
%
%
%
%
%
%
%
%
%
%
%
%
%
%
%
%
%
%
%
%
%
%
%
%


%
To explore \emph{why} known skew is not detectable in this case,
  we examine how three factors interact:
  inference error, affecting the target audience,
    and thereby indirectly affecting the delivery audience;
  algorithmic skew, affecting the delivery audience from the true target audience;
and our requirement for statistically strong evidence.
Inference error causes a large difference between our expected target audience
  and the true target audience.
Those audiences are shown in the top row of \autoref{tab:ex_baseline_corrected},
  with the same audiences used in \autoref{tab:ex_miss_skew_corrected}.
Although we expect an audience that is 15,000:15,000 Black:White (50/50\%)
  (the middle column, inferred race),
  we get an audience that is 9,621:19,460 Black:White (33/67\%), a very different ratio.

Second, we see that algorithmic skew 
  alters this ratio for delivery audience of ad 1 in \autoref{tab:ex_miss_skew_corrected}.
The true outcome is 544:1,429 Black:White (\autoref{tab:ex_miss_skew_corrected}, rightmost cell),
  but we \emph{think} it is 976:1,065 Black:White (the middle cell).
These ratios are \emph{very} different, with inferred race telling us about an even split,
  while the truth is visibly uneven.
However, the ratio of Black:White in one ad is not evidence of skew,
  instead we need to look at the relative difference of the ratios between the pair of ads.

Finally, we look at that statistical comparison.
With inferred race, that comparison is not statistically different (the bottom center cell of \autoref{tab:ex_miss_skew_corrected}).
If we knew the truth, we \emph{would} see a statistically significant difference when
  we compare delivery of ads 1 and 2, as shown in bottom right cell of \autoref{tab:ex_miss_skew_corrected}.
Exactly how these three factors interact is a function of the particular parameters
  we chose (in \autoref{sec:validation_summary} we explore many parameters to show
  which yield different outcomes).
Our point is that \emph{inference error can be an important factor}
  in auditing algorithms for delivery skew when true demographics are not known.

This second example again shows that
  \emph{inference of the target audience
  propagates to change ad impressions
  and can skew statistics about presence of skew}.
In this example,  \emph{inference of attributes in
  \textbf{does change} our conclusion}.

These examples are thought experiments
  done with a simplified model of  error
  and algorithmic skew,
  but they support our claim:
  \emph{inference error needs to be considered
  if one is to make statistically strong statements about
  presence or absence of algorithmic bias.}
  
The practical correction we propose in  \autoref{sec:new_skew_metric}
  matches the omniscient correction
  we derived using omniscient knowledge of true
  race (right-most column).
In other words,
  our solution's estimate of the number of true Black and White
  individuals that saw the first ad ($n_{1,a,f}$ and $n_{1,b,f}$)
  matches the number we would calculate using omniscient knowledge of both
  the true and inferred race ($n_{1,a,c}$ and $n_{1,b,c}$).
We prove this claim in \autoref{sec:proof_matches_omni}.

\section{Solving for $R$ and $S$}
 \label{sec:solve_R_and_S_app}
In this section,
  we prove \autoref{thm:solve_for_S}.
The proof shows we can solve for the parameters we use to model
  skew in the algorithm ($S$) and delivery rate ($R$ based on the data
  we get from the platform.
 
\begin{proof}
We know from the platform's report
  the number of inferred Black
  ($n_{1,a,i}$) and White ($n_{1,b,i}$)
  individuals that saw the ad.
We can write these quantities
 use the true racial composition of the targeted audience,
  the delivery rate ($R$) and the skew in the ad delivery algorithm $S$:

\begin{equation}\label{eq:del_aud_size_solve_S}  
\begin{aligned}
  n_{1,a,i} & = u_{a,i} (1-\myV{FDR}_{*,a}) \times R \times S  + u_{a,i} \times \myV{FDR}_{b,a} \times R \times (2-S) \\
  & + u_{a,i} \times \myV{FDR}_{o,a} \times R \times (2-S)
\end{aligned}
\end{equation}
\begin{equation*}
\begin{aligned}
n_{1,b,i} &= u_{b,i} \times \myV{FDR}_{a,b} \times R \times S  + u_{b,i} \times (1-\myV{FDR}_{*,b}) \times R \times (2-S)  \\
&+ u_{b,i} \times \myV{FDR}_{o,b} \times R \times (2-S)
\end{aligned}
\end{equation*}

\noindent We now have the above two equations with two unknowns ($R$ and $S$).
We rewrite the equations as:
$
 M \times R \times S+ N\times R - X = 0 \text{ and }
 P \times R\times S+ Q \times R - Y = 0,
$
where $|U|$ is the size of the targeted audience and we define:
$M = \frac{|U|}{2} - (|U| \times \myV{FDR}_{*,a})$, 
$N = |U| \times \myV{FDR}_{*,a}$,
$X =  n_{1,a,i}$, 
$P = |U| \times \myV{FDR}_{a,b} - \frac{|U|}{2}$,
$Q = |U| - |U| \times \myV{FDR}_{a,b}$, and
$Y = n_{1,b,i}$.
We can then a closed-form solution for $R$ and $S$
  by simply plugging in one of the equations
  into the other:
$$
R = \frac{XP - MY}{NP-MQ} \hspace{2em}
S = \frac{XQ-NY}{MY-XP}
$$
\end{proof}

Once we solve for $R$ and $S$ using this closed-form solution,
  we can estimate
  the expected inference error for the delivery
  audience of an ad.
\autoref{eq:bisg_out_error} showed one example but
  we define all error rates below:
  
\begin{equation}\label{eq:bisg_out_error_complete}
\begin{aligned}
\myV{fdr}_{b,a,1} &= \frac{u_{a,i} \times \myV{FDR}_{b,a} \times R \times (2-S)}{n_{1,a,i}} \\
\myV{fdr}_{o,a,1} &= \frac{u_{a,i} \times \myV{FDR}_{o,a} \times R \times (2-S)}{n_{1,a,i}} \\
\myV{fdr}_{a,b,1} &= \frac{u_{b,i} \times \myV{FDR}_{a,b} \times R \times (S)}{n_{1,b,i}} \\
\myV{fdr}_{o,b,1} &= \frac{u_{b,i} \times  \myV{FDR}_{o,b} \times R \times (2-S)}{n_{1,b,i}}, \\
\end{aligned}
\end{equation}

\section{Comparison of Practical and Omniscient Correction}
  \label{sec:proof_matches_omni}

In \autoref{sec:our_correction},
  we claimed that our proposed solution for correcting
  for expected inference error matches
  the correction we would apply if we had omniscient
  knowledge of both true and inferred race.
Mathematically, our claim is the following four equalities hold: $n_{1,a,f} = n_{1,a,c}$, $n_{1,b,f} = n_{1,b,c}$, $n_{2,a,f} = n_{2,a,c}$, and $n_{2,b,f} = n_{2,b,c}$.

\begin{proof}
We prove $n_{1,a,f} = n_{1,a,c}$. The other three equation can be proved similarly.

We start with the expression for $n_{1,a,f}$ we derived in \autoref{eq:corrected_stats} and plug in the expressions from \autoref{eq:bisg_out_error} for
the expected inference error in the delivery audience.

\begin{equation}\label{eq:proof_step1}
\begin{aligned}
n_{1,a,f} &= n_{1,a,i} * (1 - \myV{fdr}_{*,a,1}) + n_{1,b,i} \times \myV{fdr}_{a,b,1} \\
&= n_{1,a,i} \times \left(1 - \left(\frac{u_{a,i} \times R \times (2-S) \times \myV{FDR}_{*,a}}{n_{1,a,i}} \right)\right) \\
&+ n_{1,b,i} \times \left(\frac{u_{b,i} \times \myV{FDR}_{a,b} \times R \times S}{n_{1,b,i}}\right) \\
&= \left(n_{1,a,i} - u_{a,i} \times R \times (2-S) \times \myV{FDR}_{*,a}\right) \\
&+ \left(u_{b,i} \times \myV{FDR}_{a,b} \times R \times S\right)
\end{aligned}
\end{equation}

By rearranging the expression for $n_{1,a,i}$ from \autoref{eq:del_aud_size}, and using our knowledge that
$\myV{FDR}_{*,a} = \myV{FDR}_{b,a} + \myV{FDR}_{o,a}$, it follows that
$\left(n_{1,a,i} - u_{a,i} \times R \times (2-S) \times \myV{FDR}_{*,a}\right) = u_{a,i} \times (1 - \myV{FDR}_{*,a}) \times R \times S)$.
By substituting this expression into the last step of \autoref{eq:proof_step1}, we get:
\begin{equation}\label{eq:proof_step2}
\begin{aligned}
n_{1,a,f} &= \left(n_{1,a,i} - u_{a,i} \times R \times (2-S) \times \myV{FDR}_{*,a}\right) \\
& + \left(u_{b,i} \times \myV{FDR}_{a,b} \times R \times S\right) \\
&= (u_{a,i} \times (1 - \myV{FDR}_{*,a}) \times R \times S) + \left(u_{b,i} \times \myV{FDR}_{a,b} \times R \times S\right) \\
&= n_{1,a,c}
\end{aligned}
\end{equation}

The last step follows because
  the number of individuals inferred to be group A that saw the ad and are
  truly in group A is given by ($u_{a,i} \times (1 - \myV{FDR}_{*,a}) \times R \times S$),
  and the number of Whites individuals inferred to be in group B that saw the ad
  but are actually in group A is given by ($u_{b,i} \times \myV{FDR}_{a,b} \times R \times S$).
Taken together, the sum of the two terms gives us the
  total number of individuals that saw the ad and are truly in group A ($n_{1,a,c}$).

\end{proof}
    
\section{Varying both Inference Threshold and Audience Size}
  \label{sec:vary_both_th_aud_size}
    
In \autoref{tab:vary_th_aud},
  we provide additional simulation results where we vary both the audience size ($|U|$) and inference threshold ($\myV{Th}_{\Vs{BISG}}$)
  and check the range of values of skew in the algorithm where ignoring inference
  error leads to missing skew that exists.
Similar to the trend we observed in~\autoref{tab:vary_th_aud},
  the width of the region generally decreases
  as the audience size and BISG threshold increase.

\begin{table}
    \centering
    \begin{tabular}{c|c|c|c}
\hline
|U| & $\myV{Th}_{\Vs{BISG}}=0.5$ & $\myV{Th}_{\Vs{BISG}}=0.7$ & $\myV{Th}_{\Vs{BISG}}=0.9$   \\
\hline
10,000 	 & 0.18; [0.73, 0.91] & 0.10; [0.81, 0.91] & 0.04; [0.87, 0.91] \\
30,000 	 & 0.10; [0.85, 0.95] & 0.05; [0.90, 0.95] & 0.03; [0.92, 0.95] \\
60,000 	 & 0.08; [0.90, 0.97] & 0.05; [0.92, 0.97] & 0.03; [0.95, 0.97] \\
90,000 	 & 0.05; [0.92, 0.97] & 0.04; [0.94, 0.97] & 0.01; [0.96, 0.97] \\
120,000 	 & 0.04; [0.94, 0.97] & 0.03; [0.95, 0.97] & 0.01; [0.96, 0.97] \\
150,000 	 & 0.05; [0.94, 0.99] & 0.03; [0.96, 0.99] & 0.01; [0.97, 0.99] \\
\hline
    \end{tabular}
    \caption{Ranges of values of $S$ for which ignoring inference error leads to hiding skew that exists. For each cell, the first number indicates the width of the red shaded regions (shown in \autoref{fig:vary_all_figures_combined}), and the interval indicates the start and end values of $S$ for each region. The width of the region generally decreases as the audience size and BISG threshold increase.}
    \label{tab:vary_th_aud}
    \end{table}

\end{document}